\documentclass[11pt,letterpaper]{article}

\usepackage{epsfig}
\usepackage{amsmath}
\usepackage{amssymb}
\usepackage{amsthm}
\usepackage{indentfirst}
\usepackage{xspace}
\usepackage{multirow}
\usepackage{hyperref}
\usepackage{verbatim}
\usepackage[letterpaper,margin=1in,headheight=15pt]{geometry}
\usepackage{mdwlist}
\usepackage{framed}
\usepackage{color}
\usepackage{xcolor}
\usepackage{amscd}
\usepackage{diagrams}
\definecolor{dark-red}{rgb}{0.5,0.15,0.15}
\definecolor{dark-blue}{rgb}{0.15,0.15,0.5}
\definecolor{medium-blue}{rgb}{0,0,0.5}
\hypersetup{
    colorlinks, linkcolor={dark-red},
    citecolor={dark-blue}, urlcolor={medium-blue}
}

\definecolor{shadecolor}{rgb}{0.85,0.85,0.85}

\numberwithin{equation}{section}

\newcommand{\hx}{{\hat{x}}}
\newcommand{\hy}{{\hat{y}}}

\newcommand{\cV}{\ensuremath{\mathcal V}}
\newcommand{\cR}{\ensuremath{\mathcal R}}
\newcommand{\cB}{\ensuremath{\mathcal B}}

\newcommand{\cK}{\ensuremath{\mathcal K}}

\newcommand{\cM}{\ensuremath{\mathcal M}}
\newcommand{\cQ}{\ensuremath{\mathcal Q}}
\newcommand{\cN}{\ensuremath{\mathcal N}}
\newcommand{\cO}{\ensuremath{\mathcal O}}

\newcommand{\cY}{\ensuremath{\mathcal Y}}

\newcommand{\R}{\ensuremath{\mathbb R}}
\newcommand{\C}{\ensuremath{\mathbb C}}

\newcommand{\PP}{\ensuremath{\mathbb P}}
\newcommand{\Z}{\ensuremath{\mathbb Z}}

\newcommand{\bE}{\ensuremath{\mathbf E}}

\newcommand{\NUT}{\mathrm{NUT}}
\newcommand{\CS}{\mathrm{CS}}

\newcommand{\half}{\ensuremath{\frac{1}{2}}}

\newcommand{\N}{{\mathcal N}}

\newcommand{\kahler}{K\"ahler\xspace}
\newcommand{\hk}{hyperk\"ahler\xspace}
\newcommand{\qk}{quaternionic-K\"ahler\xspace}

\newcommand{\I}{{\mathrm i}}
\newcommand{\m}{{\mathrm m}}
\newcommand{\e}{{\mathrm e}}
\newcommand{\de}{\mathrm{d}}

\newcommand{\sing}{{\mathrm{sing}}}
\newcommand{\inst}{{\mathrm{inst}}}

\newcommand{\abs}[1]{\lvert#1\rvert}
\newcommand{\norm}[1]{\lVert#1\rVert}
\newcommand{\IP}[1]{\langle#1\rangle}

\newcommand{\ti}[1]{\textit{#1}}

\DeclareMathOperator{\Hom}{Hom}

\DeclareMathOperator{\sgn}{sgn}
\DeclareMathOperator{\Li}{Li}

\renewcommand{\sf}{\mathrm{sf}}

  {\begin{list}{}{%
    \settowidth{\labelwidth}{\textbf{#1:}}%
    \setlength{\leftmargin}{\labelwidth}\addtolength{\leftmargin}{\labelsep}}}%
  {\end{list}}

\begin{document}

\bibliographystyle{utphys}
\setcounter{page}{1}

\title{On a hyperholomorphic line bundle over the Coulomb branch}
\author{Andrew Neitzke \\ \\
Department of Mathematics \\
University of Texas at Austin}

\maketitle

{\abstract Given an $\N=2$ supersymmetric field theory in four dimensions, its dimensional
reduction on $S^1$ is a sigma model with \hk target space $\cM$.  We describe a
canonical line bundle $V$ on $\cM$, equipped with a hyperholomorphic
connection.  The construction of this connection
is similar to the known construction of the metric on $\cM$ itself:  one begins with a simple
``semiflat'' connection and then improves it by including contributions weighed by
the degeneracies of BPS particles
going around $S^1$.  We conjecture that $V$ describes the physics of the theory
dimensionally reduced on NUT space.}

\section{Introduction}

Fix a 4-dimensional $\N=2$ supersymmetric field theory $T_4$.  Upon compactification on $S^1$
of radius $R$,
one obtains a theory $T_3[R]$ whose IR description is a sigma model into a space $\cM$ with a Riemannian metric $g$.
$\cM$ is a torus bundle over the Coulomb branch $\cB$ of the theory $T_4$.

The physics of $T_4$ induces various additional structures on $\cM$.  For example,
supersymmetry implies that the metric $g$ on $\cM$ is actually \ti{hyperk\"ahler}
\cite{AlvarezGaume:1981hm}.  This metric has been considered
in some detail in various works, e.g. \cite{Seiberg:1996nz,Ooguri:1996me,Seiberg:1996ns,Gaiotto:2008cd}.
Studying the vacuum expectation values of supersymmetric line
defects of $T_4$ shows that
$\cM$ carries a \ti{canonical algebra of global holomorphic functions}, described in \cite{Gaiotto:2010be}.
(Such algebras of line operators have been discussed in connection
with topological twists of gauge theories, e.g. in \cite{Kapustin:2006hi,Kapustin:2007wm},
as well as in the context of the AGT correspondence \cite{Alday:2009aq,Alday:2009fs}.
Very similar algebras (presumably the same ones) have
also appeared in the mathematics literature:  the papers whose perspective is closest to
ours are \cite{MR2233852,MR2567745}.)
Finally, if $T_4$ has a supersymmetric surface defect with say $n$ vacua
\cite{Alday:2009fs,Gaiotto:2009fs,Taki:2010bj,Awata:2010bz}, then $\cM$ carries
a corresponding rank-$n$ \ti{hyperholomorphic vector bundle}, as discussed in \cite{Gaiotto:2011tf}.

In this note I describe yet another extra structure that $\cM$ has:  it carries a
\ti{hyperholomorphic line bundle $V$}.  (See below or Section
\ref{sec:def-hyperhol} for the meaning of this term.
In case $\cM$ has dimension $4$, a hyperholomorphic line bundle
is a \ti{$U(1)$ instanton}
on $\cM$.  More generally, a hyperholomorphic line bundle is a solution of partial
differential equations on $\cM$ which generalize the abelian self-dual Yang-Mills equations on flat space.)  While hyperholomorphic bundles also appeared in \cite{Gaiotto:2011tf}, the one we
consider here is different:  it has nontrivial topology on the torus fibers of $\cM$,
while the ones in \cite{Gaiotto:2011tf} were trivial on the fibers.  Moreover it is defined
using only the data of $T_4$ itself, with no additional structure such as a choice of surface defect.

Since $V$ is a canonical and geometrically natural object, constructed using only
the data of the $\N=2$ theory $T_4$, it seems reasonable to ask what
its physical significance is.
In Section \ref{sec:physical-interpretation} of this note
I suggest one possible physical interpretation
of $V$. I propose that when the theory $T_4$ is compactified on a Taub-NUT space of radius $R$,
the NUT center can be viewed as a local operator in theory $T_3[R]$,
and that this local operator is represented
in the 3-dimensional sigma model by a holomorphic section $s_\NUT$ of $V$.
If this proposal is correct, it would be interesting to determine $s_\NUT$ in some explicit fashion.

In the rest of the introduction I briefly summarize how $V$ is constructed, but let
me at once admit that the story as told here suffers several deficiencies.  First,
I restrict for technical convenience to the case where the theory $T_4$
has no continuous flavor symmetries at a generic point of its Coulomb branch.
(So $T_4$ could be the pure $SU(2)$
gauge theory, say, but not the theory with matter hypermultiplets.)
Second, the general case of the construction
depends on some identities for the dilogarithm function,
and I do not give a complete proof of these, only
an indication of how they ought to be proven.

\subsection*{Constructing $(V,D)$, in brief}

The bulk of this note is devoted to the construction of the bundle $V$
and its hyperholomorphic connection $D$.  This construction follows
rather similar lines to the construction of $\cM$ itself and its \hk metric $g$, given in \cite{Gaiotto:2008cd}.

We first construct $V$ as a $C^\infty$ line bundle, i.e. we determine what $V$ is \ti{topologically}.
This is already a bit nontrivial, thanks to some subtle signs which have to be taken care of,
and is described in Section \ref{sec:complex-line-bundle}.
The construction also equips $V$ with a natural Hermitian metric.
A similar construction has appeared in \cite{Alexandrov:2010np}, in a closely related context.

The next step is to equip $V$ with a hyperholomorphic structure.
$\cM$, being a \hk manifold, comes with a family of
complex structures $J_\zeta$ parameterized by $\zeta \in \C\PP^1$.  We thus have
a $\zeta$-dependent Hodge decomposition of the space of 2-forms on $\cM$.
A hyperholomorphic structure on $V$ is a unitary connection $D$, such
that for \ti{all} $\zeta$, the curvature $F_D$ is a 2-form of type $(1,1)_\zeta$.

In order to construct such a $D$, though, it is convenient to think about it in a different way.
Given $D$, we can take its $(0,1)_\zeta$ part; in this way
we obtain a family of $\bar\partial$ operators acting on sections of $V$, parameterized by $\zeta \in \C\PP^1$.
Conversely, given this family of $\bar\partial$ operators, we can recover $D$.
What we will do, roughly speaking, is
to specify such $\bar\partial$ operators ``directly'' by specifying local
holomorphic sections.  More precisely we will specify local holomorphic sections of $V$
for all $\zeta \in \C^\times$, not for all $\zeta \in \C\PP^1$; we thus obtain initially a family
of $\bar\partial$ operators parameterized by $\zeta \in \C^\times$;
but we will control the $\zeta \to 0$ and $\zeta \to \infty$
asymptotics well enough to ensure that this family of $\bar\partial$ operators extends over
$\zeta = 0$ and $\zeta = \infty$.
We can then
recover from them the desired connection $D$.  We lay out this general strategy more precisely in
Section \ref{sec:hyperhol-structures}.
It is closely related to Ward's twistorial construction of instantons \cite{Ward:1977ta}, placed in the more general context of twistor spaces of \hk manifolds \cite{Hitchin:1986ea,MR1919716}.

With this strategy understood, the main question becomes:  how to produce the desired local holomorphic sections of $V$?
Our approach is an extension of the one used in \cite{Gaiotto:2008cd}
to study the \hk metric in $\cM$ itself.  There the idea was that
the metric is determined once we know local holomorphic Darboux coordinates on $\cM_\zeta$.
These coordinates in turn were obtained as solutions of a certain integral equation,
\eqref{eq:X-integral-mult-explicit} below.\footnote{This
equation was deduced rather indirectly in \cite{Gaiotto:2008cd}; in hindsight it could have come
from a study of vacuum expectation values of supersymmetric line operators \cite{Gaiotto:2010be}.}
The basic ingredients in this equation are the Seiberg-Witten solution of the theory $T_4$ (in particular the central charge functions)
and the BPS degeneracies of $T_4$.
Our holomorphic sections are obtained in a similar way:  indeed they are determined by an integral
formula, \eqref{eq:psiinst-pieces}-\eqref{eq:psiinst2} below, which involves just the same ingredients.

We illustrate our strategy with several sorts of example.  The most basic situation is one where we neglect the
quantum corrections due to BPS particles of $T_4$, both in the construction of the \hk metric on $\cM$ and in that of the
hyperholomorphic connection $D$.  In this case all of the objects we discuss can be described by simple and explicit formulas.
We do this in Section \ref{sec:semiflat}.

The next simplest situation arises when we study theories where $T_4$ does have BPS particles, but they are all mutually local with one another
(i.e. there is a duality frame in which they are all electrically charged).  In this case one can compute the full \hk metric
in $\cM$ explicitly, summing up all quantum corrections \cite{Ooguri:1996me,Seiberg:1996ns,Gaiotto:2008cd}.
In Section \ref{sec:ov} we study the simplest example,
the $U(1)$ theory with a single charge-1 hypermultiplet.  Here the metric on $\cM$ is the famous ``Ooguri-Vafa metric.''
Our construction of the hyperholomorphic
connection $D$ turns out to be similarly explicit in this case.  In particular, we can see that the quantum corrections from the BPS
instantons serve not only to smooth out the metric on $\cM$ (which would naively be singular at the locus where a BPS particle
becomes massless) but also to smooth out $D$.

Finally, in Section \ref{sec:general} we confront the most general situation.  Here neither
the metric in $\cM$ or the hyperholomorphic connection $D$ can be described in a simple closed
form.  Moreover, an important new feature appears:  BPS bound states can appear or decay at the walls of marginal stability
in $\cB$.  This is the ``wall-crossing'' phenomenon
studied e.g. in \cite{Cecotti:1993rm,Seiberg:1994rs,Denef:2007vg,ks1,Gaiotto:2008cd,Cecotti:2009uf,Cecotti:2010qn,Manschot:2010qz}.
As a result of this phenomenon the smoothness of $D$ is not
\ti{a priori} obvious from the construction:  one could fear that $D$ would be discontinuous along the walls of marginal stability.
For the metric on $\cM$, a similar issue arose in \cite{Gaiotto:2008cd}.  In that case
it turned out that the Kontsevich-Soibelman
wall-crossing formula for the BPS degeneracies \cite{ks1} was sufficient to ensure smoothness.
In the present case, on the other hand,
the necessary smoothness seems to be equivalent to
certain dilogarithm identities, which are not obviously reducible to the Kontsevich-Soibelman
formula.  It is natural to expect that the needed identities
follow from the ``refined wall-crossing formula'' for the BPS degeneracies, discussed e.g. in \cite{Dimofte:2009tm,Cecotti:2009uf,Gaiotto:2010be}.
Unfortunately, in this note I give only a brief sketch of how this ought to be proven,
except in the simplest example, where the needed identity is just
the well-known ``pentagon identity'' of the dilogarithm.

\subsection*{Closely related work}

The idea that $\cM$
might carry an interesting hyperholomorphic line bundle goes back to unpublished joint work with
Boris Pioline in 2008.
Our interest at that time was in $\N=2$, $d=4$ supergravity.  Compactifying such a supergravity theory on $S^1$ leads to a 3-dimensional theory for which the vector multiplet
moduli space is a \qk manifold $\cQ$.  On the other hand, such a supergravity theory can often
be described by coupling a rigid, superconformally invariant $\N=2$ field theory $T_4$
to a conformal supergravity multiplet.
One could then obtain a pseudo-\hk manifold $\cM$ by compactifying $T_4$ on $S^1$.\footnote{The theory $T_4$ in this case has an indefinite
kinetic term, which explains why $\cM$ is pseudo-\hk rather than \hk.}
At least when quantum corrections are neglected, we noticed that
$\cM$ can be thought of as a \hk quotient of
the Swann bundle over $\cQ$ by the group $\R^\times$.
From this picture it follows that $\cM$ should carry a natural hyperholomorphic line
bundle, and we wrote an explicit formula for its curvature, essentially \eqref{eq:hh-curv-sf} below.

More recently, I and Alexandrov-Persson-Pioline independently realized that this hyperholomorphic line bundle continues to exist
when quantum corrections to the metric of $\cM$ are included; this is an important special case of the construction described in this note.
This note thus has considerable overlap with the paper \cite{Alexandrov:2011ac}
of Alexandrov-Persson-Pioline.  (I thank the authors for sharing an advance draft of
\cite{Alexandrov:2011ac}.)

Also while this note was in preparation, I learned from Tamas Hausel about some independent
work of Haydys \cite{MR2394039} which is also related to the above special case, as follows.
The $\cM$ considered in the preceding paragraph are special in that they have a $U(1)$ isometry (following from the unbroken $R$-symmetry,
in turn following from the superconformal invariance).  This $U(1)$ isometry preserves one
of the complex structures of $\cM$ while rotating the others
into one another.  Haydys's work
implies \cite{hitchin-haydys-talk} that for \ti{any} \hk manifold $\cM$ with such a $U(1)$ isometry,
(acting in a Hamiltonian fashion with respect to the preserved \kahler form and obeying an appropriate quantization condition),
there is a natural hyperholomorphic line bundle.
For the $\cM$ of the preceding paragraph, this line bundle should coincide
with our $V$.
Haydys also discussed a corresponding
\qk space $\cQ$, whose relation to $\cM$ is just as described above.

\section{Review} \label{sec:review}

\subsection{IR data in $d=4$}

We will be considering quantum systems with 4-dimensional Poincare invariance
and $\N=2$ supersymmetry.
The IR physics of such a system,
in a Coulomb phase, is described by \ti{abelian} $\N=2$ gauge theory.  Such a theory
is determined (to lowest order in the derivative expansion) by a short list of ``holomorphic'' data,
as described by Seiberg and Witten \cite{Seiberg:1994rs,Seiberg:1994aj}.  Here we briefly review that data and fix notation; for more details see \cite{Gaiotto:2008cd}.

\begin{itemize}
 \item \ti{Coulomb branch}:  a complex manifold $\cB$.
Parameterized by vector multiplet scalars, $\cB$
is (one component of) the exact moduli space of quantum vacua of the system.
In the most familiar examples, $\cB$ is actually a complex affine space.
 \item \ti{Discriminant locus}:  a divisor $\cB_\sing \subset \cB$.  $\cB_\sing$ consists of the vacua where the description
of the physics by \ti{pure} abelian gauge theory breaks down.  We also define $\cB^* = \cB \setminus \cB_\sing$.
 \item \ti{Charge lattice}:  a local system of lattices $\Gamma$ over $\cB^*$.  Around each point
of $\cB^*$ we have a local description of the physics by a gauge theory with gauge group isomorphic to
$U(1)^{r}$; the fiber $(\Gamma)_u$ of $\Gamma$ over $u$
is the lattice of electromagnetic charges in that theory.
It has rank $2 r$,
and is equipped with the antisymmetric $\Z$-valued DSZ pairing $\IP{\cdot,\cdot}$.
Throughout this paper, we make the assumption that this pairing has determinant $1$,
i.e. it establishes an isomorphism $\Gamma \simeq \Gamma^*$.
We will abuse notation by writing ``$\gamma \in \Gamma$'' to refer
to a locally constant section of $\Gamma$.  Such sections generally do not exist globally, so any formula in which
the symbol $\gamma$ appears must always be interpreted as making sense \ti{locally} over a patch of $\cB^*$.
 \item \ti{Central charge}:  a homomorphism $Z: \Gamma \to \C$, varying holomorphically over $\cB^*$.
Concretely this means that for any $\gamma \in \Gamma$, defined over some patch $U \subset \cB^*$,
we have a holomorphic function $Z_\gamma: U \to \C$, obeying $Z_{\gamma + \gamma'} = Z_\gamma + Z_{\gamma'}$.
\end{itemize}

These data are subject to some constraints.  In particular, the central charges $Z$ are not arbitrary
but rather must be derivable from some prepotential; this imposes the condition
\begin{equation} \label{eq:lagrangian-condition}
 \IP{\de Z, \de Z} = 0.
\end{equation}
Moreover, the kinetic energy of the scalar fields should be positive definite, which requires that
$\IP{\de Z, \de \bar{Z}}$ is a \ti{positive} 2-form.

Our assumptions above are less general than what is usually considered, in that we have assumed that there
are no continuous flavor symmetries in the theory.  (So we could be considering the pure $SU(2)$
gauge theory, say, but not the theory with matter hypermultiplets.)

\subsection{Compactification to $d=3$} \label{sec:compactification}

Now we consider the theory compactified from $d=4$ to $d=3$ on a circle of circumference $2 \pi R$,
in a Coulomb phase, in the IR.  As described in \cite{Seiberg:1996nz},
the resulting theory is a sigma model into a \hk manifold $\cM$.

Topologically
$\cM$ is fibered over $\cB$, $\pi: \cM \to \cB$, since the scalar fields of the $d=4$ theory
survive into the $d=3$ compactification.
Let $\cM^* = \pi^{-1}(\cB^*)$.
For any
$u \in \cB^*$, the fiber $\cM_u = \pi^{-1}(u)$ is a torus, parameterized by the Wilson lines of the $r$ abelian gauge
fields around the compactification circle, and their $r$ magnetic duals.
A more duality-invariant way of saying this is that the torus fiber
$\cM_u$ admits functions $\varphi_\gamma: \cM_u \to U(1)$ for charges $\gamma \in \Gamma$.
Choosing any basis of $\Gamma$ we would obtain a set of circle-valued coordinates $\varphi_{\gamma_i}$, $i = 1, \dots, 2r$.
The functions $\varphi_\gamma$ obey a twisted additive law,\footnote{The physical origin of this twisting is still not completely clear, but see
previous discussion in \cite{Gaiotto:2008cd,Gaiotto:2010be}.}
\begin{equation} \label{eq:theta-twisted-additive}
 \varphi_\gamma \varphi_{\gamma'} = (-1)^{\IP{\gamma, \gamma'}} \varphi_{\gamma + \gamma'}.
\end{equation}

$\N=2$ supersymmetry of the IR
description dictates that in $d=4$ we have one complex scalar for each abelian gauge field,
so in particular, $\cB$ has complex dimension $r$.
The torus fiber $\cM_u$ and the base $\cB$ thus have the same real dimension $2 r$.

The fibers of $\cM$ above $\cB_\sing \subset \cB$ are trickier and
cannot be determined using abelian gauge theory alone;
in general they are some kind of degenerations of tori.

\subsection{Basis functions $\cY_\gamma$} \label{sec:basis-functions}

As $\cM$ is a \hk manifold, it admits a distinguished family of complex structures $J_{\zeta}$, parameterized by $\zeta \in \C\PP^1$.  We write $\cM_\zeta$ for the complex manifold $(\cM, J_\zeta)$.
This family of complex structures corresponds to a family of subalgebras $A_\zeta$ of the $\N=2$ algebra.

The vacuum expectation value of any operator in the compactified theory is a function on $\cM$.
Vacuum expectation values of $A_\zeta$-invariant operators are holomorphic functions on $\cM_\zeta$.
For $\zeta \in \C^\times$, $A_\zeta$ contains the translations along the compactification circle (and no others),
so there are no $A_\zeta$-invariant local operators, but
there are $A_\zeta$-invariant line operators wrapped around the circle.
In \cite{Gaiotto:2008cd,Gaiotto:2009hg,Gaiotto:2010be}
it was shown that the vacuum expectation values of such line operators
can be naturally expressed in terms of more ``elementary'' holomorphic functions $\cY_\gamma$.
(Morally one would like to think of $\cY_\gamma$ as vacuum expectation values of ``IR line operators,''
and this decomposition as expressing the way the RG flow relates UV line operators to IR ones,
but this is a tricky notion to make precise.)

Each $\cY_\gamma$ is a function on $\cM \times \C^\times$, which for any fixed $\zeta \in \C^\times$ is holomorphic
on $\cM_\zeta$.  They obey a twisted multiplicative law,
\begin{equation}
 \cY_\gamma \cY_{\gamma'} = (-1)^{\IP{\gamma,\gamma'}} \cY_{\gamma+\gamma'}.
\end{equation}
A first approximation to these functions is provided by the explicit formula
\begin{equation} \label{eq:y-sf}
 \cY_\gamma^\sf(\zeta) = \exp \left[  \pi R (\zeta^{-1} Z_\gamma + \zeta \bar{Z}_\gamma) \right] \varphi_\gamma.
\end{equation}
This approximation becomes exact in the limit $R \to \infty$.
The corrections are
controlled by the 4d BPS degeneracies (second helicity supertrace) $\Omega(\gamma) \in \Z$.
At least when the radius $R$ is sufficiently large, the corrected $\cY_\gamma$ can be
characterized as solutions to an integral equation \cite{Gaiotto:2008cd}:
\begin{equation} \label{eq:X-integral-mult-explicit}
\cY_\gamma(\zeta) = \cY^\sf_\gamma(\zeta) \exp \left[ -\frac{1}{4
\pi \I} \sum_{\gamma'} \Omega(\gamma') \langle \gamma,\gamma'
\rangle \int_{\ell_{\gamma'}} \frac{\de \zeta'}{\zeta'} \frac{\zeta' +
\zeta}{\zeta' - \zeta} \log (1-\cY_{\gamma'}(\zeta'))\right],
\end{equation}
where the contour of integration is the ``BPS ray''
\begin{equation}
 \ell_{\gamma'} = Z_{\gamma'} \R_-.
\end{equation}
At least for sufficiently large $R$, $\abs{\cY_{\gamma'}(\zeta')} < 1$ when $\zeta' \in \ell_{\gamma'}$; so we can define the $\log(1-\cY_{\gamma'}(\zeta'))$
appearing on the right of \eqref{eq:X-integral-mult-explicit} unambiguously
to be the \ti{principal} branch.

An immediate consequence of this equation is that the functions $\cY_\gamma$ are actually \ti{discontinuous}.
The discontinuity appears precisely when $\zeta$ lies on the ray $\ell_{\gamma'}$.
At this locus the integral in \eqref{eq:X-integral-mult-explicit} is ill-defined because the denominator
$\zeta' - \zeta$ vanishes.  The discontinuity is easily evaluated by the residue formula.  Let $\cY_\gamma^\pm$ denote
the limits of $\cY_\gamma$ as we approach $\ell_{\gamma'}$ from both sides: then we have \begin{equation} \label{eq:y-disc}
 \cY^+_\gamma = \cY^-_\gamma (1 - \cY_{\gamma'})^{\Omega(\gamma') \IP{\gamma,\gamma'}}.
\end{equation}

The $\cY_\gamma$
are local Darboux coordinates for the holomorphic symplectic form on $\cM_\zeta$:
\begin{equation}
 \varpi(\zeta) = \frac{1}{4 \pi^2 R} \IP{\de \log \cY, \de \log \cY}.
\end{equation}
(Although the individual functions $\cY_\gamma$ are discontinuous, $\varpi(\zeta)$ is continuous,
as it must be.)
In particular, this is enough information to determine the full \hk structure on $\cM$.

\subsection{A geometric view} \label{sec:geometric}

We have just reviewed a description of the \hk structure on $\cM$ in terms of some
locally defined basis functions $\cY_\gamma$.
For use in the rest of this note, it will be useful to have a slightly more geometric
point of view on this construction.

We consider an
auxiliary complex torus $T$ with coordinate functions $Y_\gamma: T \to \C^\times$ for
$\gamma \in \Gamma$, obeying the relation\footnote{As usual, this is a local construction over a patch of
$\cB$ where $\Gamma$ can be trivialized.  It could be made global by considering $T$ to be a \ti{local system}
of tori.  In that case the map $\cY$ becomes a globally defined object, although the individual $\cY_\gamma$ are
still only local.
This is the most perfectly invariant description of the situation, but hard experience has shown that it
also tends to be perfectly confusing to the reader.}
\begin{equation}
 Y_\gamma Y_{\gamma'} = (-1)^{\IP{\gamma,\gamma'}} Y_{\gamma+\gamma'}.
\end{equation}
$T$ carries a canonical holomorphic symplectic form induced by the pairing $\IP{,}$, namely
$\omega_T = \IP{ \de \log Y, \de \log Y }$.

Then we can think of the functions $\cY_\gamma$, $\gamma \in \Gamma$,
as the components of a single map $\cY: \cM \to T$:  equivalently,
\begin{equation}
 \cY_\gamma = \cY^* Y_\gamma.
\end{equation}
Up to some irrelevant overall factors, $\varpi$ is the pullback to $\cM$ of a canonical
complex symplectic form on $T$:
\begin{equation}
 \varpi = \frac{1}{4 \pi^2 R} \cY^* \omega_T.
\end{equation}
In other words, the construction of the complex symplectic structure $\varpi$ on
$\cM$ is simply to pull back that structure from the complex symplectic torus $T$, via
the map $\cY$.

For any $\gamma' \in \Gamma$, we can define a (birational) automorphism
$\cK_{\gamma'}$ of $T$ by
\begin{equation}
 \cK_{\gamma'}^* X_{\gamma} = X_\gamma ( 1 - X_{\gamma'})^{\IP{\gamma,\gamma'}}.
\end{equation}
(These automorphisms appeared prominently in \cite{ks1}.)
The discontinuities of the functions $\cY_\gamma$ then can be expressed in terms of the map
$\cY$ as
\begin{equation}
 \cY^+ = \cK_{\gamma'}^{\Omega(\gamma')} \circ \cY^-.
\end{equation}
So as we cross the ray $\ell_{\gamma'}$, the map $\cY$ jumps by composition with the automorphism $\cK_{\gamma'}^{\Omega(\gamma')}$.

The torus $T$ possesses a natural real structure map, $\rho: T \to \bar{T}$, defined by
\begin{equation}
 \rho^* Y_\gamma = \overline{Y}_{-\gamma}.
\end{equation}
The fixed locus of $\rho$ is the locus where all $\abs{Y_\gamma} = 1$, a real torus.
The map $\cY$ obeys the reality condition
\begin{equation}
 \cY(\zeta) = \rho^* \cY(-1/\bar\zeta).
\end{equation}

\section{A complex line bundle $V$ over $\cM$} \label{sec:complex-line-bundle}

The purpose of this section is to describe a complex line bundle $V$ over $\cM$, equipped with a canonical Hermitian metric $\norm{\cdot}$
and unitary connection $\nabla$.

Restricted to each torus fiber $\cM_u$, the construction of $V$ is basically a standard one;
the only part which may be slightly novel
is the role of the twisting function, which serves to make
$V$ completely canonical (rather than depending on a choice of quadratic refinement
of the pairing $\IP{,}$).  One has to extend this standard construction from a single fiber $\cM_u$
to the whole of $\cM$.
This is done using the flat parallel transport given by the canonical
coordinates $\varphi_\gamma$.
A construction closely related to the one described here was given previously in
\cite{Alexandrov:2010np}.

We will first produce a line bundle $V^\sigma$ which depends on
a choice of a quadratic refinement $\sigma$, i.e. a map $\sigma: \Gamma \to \Z_2$ obeying
\begin{equation}
 \sigma(\gamma) \sigma(\gamma') = (-1)^{\IP{\gamma,\gamma'}} \sigma(\gamma + \gamma').
\end{equation}
The quadratic refinement $\sigma$
may exist only locally, so $V^\sigma$ generally lives only over a local patch $\cM^\sigma$ of $\cM$.
Then we will give isomorphisms $\iota^{\sigma \sigma'}: V^{\sigma'} \to V^{\sigma}$ for any $\sigma$, $\sigma'$,
obeying $\iota^{\sigma \sigma'} \circ \iota^{\sigma' \sigma''} = \iota^{\sigma \sigma''}$.
Using these to glue, we get a line bundle $V$ over the whole $\cM$.
It would be nice to have a more
direct way of producing $V$ without using the intermediaries $V^\sigma$.

So, choose a quadratic refinement $\sigma$.  We introduce new coordinate functions
on $\cM^\sigma$,
\begin{equation} \label{eq:tcoord}
 \varphi^\sigma_\gamma = \sigma(\gamma) \varphi_\gamma.
\end{equation}
These functions obey an \ti{untwisted} multiplicative law,
\begin{equation}
 \varphi^\sigma_\gamma \varphi^\sigma_{\gamma'} = \varphi^\sigma_{\gamma + \gamma'}.
\end{equation}
The $\varphi^\sigma_\gamma$ do not admit single-valued logarithms on $\cM^\sigma$; let $\cN^\sigma$ be the minimal
covering space on which all such logarithms do exist.  $\cN^\sigma$ is a fiberwise universal
cover of $\cM^\sigma$, with fiber the linear space $\Gamma^* \otimes_\Z \R$.  Given a point
$\theta^\sigma$ in this fiber we have
\begin{equation} \label{eq:log}
 \varphi^\sigma_\gamma = e^{\I \theta^\sigma_\gamma}.
\end{equation}
$\cM^\sigma$ can be recovered from $\cN^\sigma$ by dividing out by an action of $\Gamma^*$,
\begin{equation}
 T_\delta (\theta^\sigma) = \theta^\sigma + 2 \pi \delta.
\end{equation}
Now we consider the trivial line bundle $W^\sigma = \C \times \cN^\sigma$ over $\cN^\sigma$.
$W^\sigma$ carries a connection $\nabla = \de + A^\sigma$ where $A^\sigma$ is a 1-form on $\cN^\sigma$,
\begin{equation} \label{eq:A-sigma}
A^\sigma = \frac{\I}{4 \pi} \IP{\theta^\sigma, \de \theta^\sigma}.
\end{equation}
The action of $\Gamma^*$ on $\cN^\sigma$ lifts to an action on $W^\sigma$,
\begin{equation} \label{eq:gammastar-action}
 T_\delta (\theta^\sigma, \psi^\sigma) = (\theta^\sigma + 2 \pi \delta, \psi^\sigma \sigma(\delta) e^{\frac{\I}{2} \IP{\theta^\sigma, \delta}}).
\end{equation}
Here we have used the isomorphism $\Gamma \simeq \Gamma^*$ in order to define the symbol
$\sigma(\delta)$.  One checks directly that $T_\delta T_{\delta'} = T_{\delta + \delta'}$; this would have
failed without including $\sigma(\delta)$.  Moreover, the connection $\nabla$ is invariant under $T_\delta$.
We thus obtain a quotient line bundle $V^\sigma = W^\sigma / \Gamma^*$
over $\cM^\sigma$, with connection.

Finally, suppose given two quadratic refinements $\sigma$, $\sigma'$.  The two differ
by $\sigma(\gamma) / \sigma'(\gamma) = (-1)^{c^{\sigma \sigma'} \cdot \gamma}$, for some
$c^{\sigma \sigma'} \in \Gamma^*$ (in fact there are many
possible $c^{\sigma \sigma'}$ obeying this equation since we are free to shift by an element of
$2 \Gamma^*$).  This $c^{\sigma \sigma'}$ determines a map
$W^\sigma \to W^{\sigma'}$ by
\begin{align}
\theta^{\sigma'} &= \theta^{\sigma} + \pi c^{\sigma\sigma'}, \label{eq:sig-shift-real-1} \\
\psi^{\sigma'} &= \psi^\sigma e^{\frac{\I}{4} \IP{\theta^\sigma, c^{\sigma\sigma'}}}. \label{eq:sig-shift-real-2}
\end{align}
A short computation shows this map indeed commutes with the action of $\Gamma^*$
and thus descends to an isomorphism between $V^\sigma$ and $V^{\sigma'}$, which moreover
is independent of the choice of $c^{\sigma \sigma'}$, and is compatible with
the connections.  This gives the
isomorphism $\iota^{\sigma \sigma'}$ we claimed.
That completes the construction.

While the $\theta^\sigma$ do not descend to $\cM$, their differentials do; indeed this gives
a canonical $\Gamma^*$-valued 1-form $\de \theta$ on $\cM$.
The curvature of the connection $\nabla$ in $V$ is a 2-form on $\cM$,
\begin{equation} \label{eq:d-curv}
 F = \frac{\I}{4 \pi} \IP{\de \theta, \de \theta}.
\end{equation}
This formula
implies in particular that $V$ is ``minimally quantized'':  the integrals of $\frac{1}{2 \pi \I} F$
along 2-cycles in the torus fiber generate $\Z$.  (Much of the subtlety of the above construction
was related to the fact that we were trying to get this minimally quantized bundle.  If we had settled
for a construction of $V^2$ instead of $V$, the discussion would have been considerably simpler.)

\section{Constructing a hyperholomorphic structure on $V$} \label{sec:hyperhol-structures}

\subsection{Hyperholomorphic structures} \label{sec:def-hyperhol}

Our goal in the rest of the paper is to
describe a \ti{hyperholomorphic structure} on the Hermitian line bundle $V$.
In one sentence, what this means is that we will give a unitary connection $D$ in $V$
such that the curvature 2-form $F$ on $\cM$ is of type $(1,1)_\zeta$ for all $\zeta \in \C\PP^1$.

The name ``hyperholomorphic'' arises from the fundamental fact that, just as a \hk manifold $\cM$ is a \kahler manifold in a $\C\PP^1$ worth of ways,
a hyperholomorphic line bundle $V$ is a holomorphic line bundle
in a $\C\PP^1$ worth of ways.
To see this, suppose we have found a $D$
as above.
For each $\zeta$, let $D^{(0,1)_\zeta}$ denote the $(0,1)$ part of $D$, with respect to complex structure $J_\zeta$
on $\cM$.  The vanishing
of $F^{(0,2)_\zeta}$ is equivalent to the statement that $D^{(0,1)_\zeta}$ squares to zero when acting on form-valued sections, and so
defines a ``Cauchy-Riemann operator'' on $V$ over $\cM_\zeta$.  In other words, we can
equip $V$ with the structure of a holomorphic line bundle $V_\zeta$ over
$\cM_\zeta$ by declaring that the holomorphic sections of $V_\zeta$ are just the sections of $V$
that are annihilated by $D^{(0,1)_\zeta}$.

Our approach to constructing the connection $D$ will be to construct directly the
holomorphic structures $V_\zeta$ on $V$ for $\zeta \in \C^\times$, in a way that varies holomorphically
with $\zeta$, while paying enough attention to the
asymptotics to ensure that the holomorphic structure
actually extends to $\zeta = 0$ and $\zeta = \infty$.
As it turns out, this will be enough to show that these holomorphic structures
indeed come from a single connection $D$ in $V$.

\subsection{The bundle $\cV$} \label{sec:bundle-def}

We begin by
defining a holomorphic line bundle $\cV$ over the torus $T$ which appeared in Section
\ref{sec:geometric}.
The definition of $\cV$ is essentially an analytic continuation of the
definition of $V$ we gave in Section \ref{sec:complex-line-bundle}:  we just need to replace the real torus $\cM_u \simeq (S^1)^{2 r}$
by the complex torus $T \simeq (\C^\times)^{2 r}$.

Thus, following the pattern of Section \ref{sec:complex-line-bundle},
we first produce a line bundle $\cV^\sigma$ over $T$ which depends on
a choice of a quadratic refinement $\sigma$, then identify the different $\cV^\sigma$
to get a single bundle $\cV$.  I do not repeat the whole story here;
the only difference from Section \ref{sec:complex-line-bundle} is that
we replace the $U(1)$-valued functions $\varphi_\gamma$
by the $\C^\times$-valued functions $X_\gamma$,
and similarly replace their $\R$-valued logarithms $\theta^\sigma_\gamma$ by $\C$-valued logarithms $x^\sigma_\gamma$, functions on a covering $U^\sigma$ of $T$.

For future reference we write the most important formulas here.
The analogue of \eqref{eq:gammastar-action}, giving the $\Gamma^*$-action on the bundle,
is
\begin{equation} \label{eq:gammastar-action-complex}
 T_\delta (x^\sigma, \psi^\sigma) = (x^\sigma + 2 \pi \delta, \psi^\sigma \sigma(\delta) e^{\frac{\I}{2} \IP{x^\sigma, \delta}}).
\end{equation}
The analogue of \eqref{eq:sig-shift-real-1}, \eqref{eq:sig-shift-real-2} giving the effect
of changing the quadratic refinement is
\begin{align}
x^{\sigma'} &= x^{\sigma} + \pi c^{\sigma\sigma'}, \label{eq:sig-shift-complex-1} \\
\psi^{\sigma'} &= \psi^\sigma e^{\frac{\I}{4} \IP{x^\sigma, c^{\sigma\sigma'}}}. \label{eq:sig-shift-complex-2}
\end{align}

Just as $V$ carried a canonical connection, $\cV$ carries a canonical holomorphic connection.
The curvature of this holomorphic connection is the symplectic form $\omega_T$.
Unlike $V$, though, $\cV$ does not have any natural Hermitian structure; what it has
instead is a natural isomorphism
\begin{equation}
\rho^* \cV \simeq \bar{\cV}^*.
\end{equation}

In the rest of this note we will generally encounter not $\cV$ directly
but rather its pullback $\cY(\zeta)^* \cV$ to $\cM$:  our aim is to identify $V$ with this pullback and thus
induce a family of holomorphic structures $V_\zeta$ on $V$.
In thinking about this pullback two important issues arise, one
concerning logarithms, the other concerning dilogarithms.  We turn next to these points.

\subsection{Canonical logarithms}

The first issue is relatively straightforward:  it concerns the existence of the ``logarithm'' of $\cY$.
More precisely:  since both $V$ and $\cV$ are defined
using coverings of the spaces where they live, it will be technically useful to know that
the map $\cY(\zeta)$ lifts to these coverings.  We call the lifted map $\log \cY$:
\begin{equation}
\begin{CD}
 \cN^\sigma @>\log \cY(\zeta)>> U^\sigma \\
   @VVV             @VVV            \\
    \cM     @>\cY(\zeta)>>     T
\end{CD}
\end{equation}
The existence of such a lifting can easily be seen for the semiflat approximation $\cY^\sf$,
which implies it holds also for the exact $\cY$ at least when $R$ is large enough.
In particular we can pull back
the functions $x^\sigma_\gamma$ from $U^\sigma$ to $\cN^\sigma$:  we define
\begin{equation}
 \xi^\sigma_\gamma = (\log \cY)^* x^\sigma_\gamma.
\end{equation}
On $\cN^\sigma$, these pulled-back functions obey a complexified analog of \eqref{eq:tcoord}, \eqref{eq:log}:
\begin{equation}
 \cY_\gamma = \sigma(\gamma) e^{\I \xi^\sigma_\gamma}.
\end{equation}
So up to the tricky sign $\sigma(\gamma)$, $\xi_\gamma$ is a logarithm of $\cY_\gamma$.

\subsection{Canonical dilogarithm}

There is also a second, more serious difficulty in talking about $\cY(\zeta)^* \cV$.
As we have explained
in Sections \ref{sec:basis-functions}-\ref{sec:geometric}, $\cY(\zeta)$ is a \ti{discontinuous} map:  when we cross
the ray $\ell_{\gamma'}$, it jumps by composition with the automorphism $\cK_\gamma^{\Omega(\gamma)}$ of $T$.  This being so,
what should we even \ti{mean} by $\cY(\zeta)^* \cV$?

Recall that we denoted by $\cY^\pm$ the two limits of $\cY$ as $\zeta$ approaches $\ell_{\gamma'}$ from the two possible directions.
Our problem is that the two line bundles $(\cY^+)^* \cV$ and $(\cY^-)^* \cV$ appear to be different.
The latter can also be written as $(\cY^+)^* (\cK_{\gamma'}^{\Omega(\gamma')})^* \cV$.
After this rewriting, the apparent difficulty is to relate
$\cV$ and $(\cK_{\gamma'}^{\Omega(\gamma')})^* \cV$.

Luckily, these two bundles are actually canonically isomorphic.  Indeed, there is
a natural trivializing section $L_\gamma$ of the difference bundle
$\Hom(\cV, \cK_{\gamma}^* \cV)$.
Thus, what we will mean by a ``holomorphic section of $\cY(\zeta)^* \cV$''
is a holomorphic section $s$ of $\cY(\zeta)^* \cV$
away from the rays $\ell_{\gamma'}$, such that the limits
$s^+$, $s^-$ at $\ell_{\gamma'}$
are related by $s^+ = s^- \otimes (\cY^-)^* L_{\gamma'}^{\Omega(\gamma')}$.
With this notion of holomorphic section,
$\cY(\zeta)^* \cV$ becomes an honest holomorphic line bundle over $\cM$.

To see that $L_\gamma$ exists we will just write a concrete formula for it.  To do so, though, we again need to work not on $T$ but on some covering.
As we have just described,
sections of $\cV$ can be conveniently described as functions on $U^\sigma$ with an appropriate $\Gamma^*$-equivariance.
On the other hand, sections of $\cK_\gamma^* \cV$ cannot be described this way:  rather they are functions on the pulled-back covering $\cK_\gamma^* U^\sigma$.  To describe $L_\gamma$ we
work over the smallest common refinement of these two coverings:

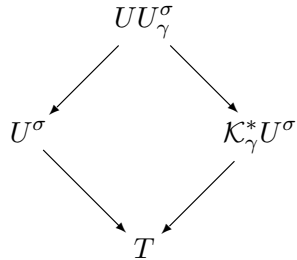
\begin{diagram}[small]
       &        & UU^\sigma_\gamma  &        &    \\
       &  \ldTo &                   & \rdTo  &    \\
 U^\sigma &        &                   &        & \cK_\gamma^* U^\sigma   \\
       & \rdTo  &                   & \ldTo  &    \\
       &        &       T        &        &    \\
\end{diagram}
On the common refinement $UU^\sigma_\gamma$, we have not only a single-valued logarithm
of $Y_\gamma$ but also of $(1-Y_\gamma)$.
We define the desired $L_\gamma^\sigma$ by
\begin{equation}
 L_\gamma^\sigma = \exp \left[ \frac{1}{2 \pi \I} (\Li_2(Y_\gamma) + \half x_\gamma^\sigma \log(1 - Y_\gamma)) \right],
\end{equation}
which is also single-valued on this covering.  One can check directly using the
branch structure of $\Li_2$ that $L_\gamma^\sigma$ is indeed a section
of $\Hom(\cV, \cK_{\gamma}^* \cV)$.

\subsection{Identifying $V$ and $\cV_\zeta$} \label{sec:identifying}

We now define a holomorphic line bundle $\cV_{\zeta}$
on $\cM_\zeta$ by
\begin{equation}
 \cV_{\zeta} = \cY(\zeta)^* \cV.
\end{equation}
$\cV_\zeta$ is topologically the same as $V$, so now we are in a position to build a family of holomorphic
structures $V_\zeta$ on $V$:  all we need is an identification
between $V$ and $\cV_\zeta$, i.e. a nowhere-vanishing
section $\Psi(\zeta)$ of the line bundle $\Hom(\cV_\zeta, V)$.
Our goal is to construct this family of holomorphic structures in such a way that the
$V_\zeta$ all come from an underlying hyperholomorphic connection $D$ in $V$.

In the next few sections we will explain how
$\Psi(\zeta)$ will be constructed.  For now we focus on what properties of $\Psi(\zeta)$ are needed in order to guarantee that the $V_\zeta$ indeed come from a hyperholomorphic connection.
We state these properties as ``assumptions'' about $\Psi$; in later sections we will see that
our assumptions indeed hold for the $\Psi(\zeta)$ we construct.

\begin{itemize}

\item It is known from \cite{Gaiotto:2008cd} that locally on $\cM$ one can find complex vector fields
$v_i^{(0)}$, $v_i^{(1)}$, $i = 1, \dots, 2 r$, such that
\begin{equation} \label{eq:01-decomp}
 v_i(\zeta) = v^{(0)}_i + \zeta v^{(1)}_i
\end{equation}
span $T^{0,1} \cM_\zeta$, and
\begin{equation} \label{eq:v-reality}
 v_{i+r}(\zeta) = \zeta \overline{v_i(-1/\bar\zeta)}.
\end{equation}
If we fix a trivialization of $V$, we may define functions $f_i(\zeta)$ on $\cM$ by
\begin{equation} \label{eq:cr-psi}
\left( \bar\partial_{v_i(\zeta)} + f_i(\zeta) \right) \Psi(\zeta) = 0 \qquad (i = 1, \dots, 2 r)
\end{equation}
where $\bar\partial$ is the Cauchy-Riemann
operator giving the holomorphic structure in $\cV_\zeta^*$.

Our first assumption on $\Psi(\zeta)$ is that $f_i(\zeta)$ has the same $\zeta$ dependence as does $v_i(\zeta)$, i.e.
\begin{equation} \label{eq:fi-decomp}
f_i(\zeta) =  f^{(0)}_i +  \zeta f^{(1)}_i.
\end{equation}
(Below we will interpret
\eqref{eq:cr-psi} as the Cauchy-Riemann equations saying that $\Psi$ is a holomorphic
section of $\Hom(\cV_\zeta, V_\zeta)$.  In particular, the coefficient functions
$f_i(\zeta)$ determine the Cauchy-Riemann operator and thus the
holomorphic structure in $V_\zeta$.)

\item For each fixed $x \in \cM$, we assume $\Psi(x; \zeta)$ is holomorphic in $\zeta$ on a dense
subset of $\C^\times$.  (In our application, $\Psi(x; \zeta)$ will be holomorphic away
from a countable union of rays running from $0$ to $\infty$.)

\item We assume $\Psi$ obeys the reality condition
\begin{equation} \label{eq:psi-reality}
\Psi(\zeta) \overline{\Psi(-1/\bar\zeta)} = 1.
\end{equation}
(In writing this equation we used
the isomorphisms $\rho^* \cV \simeq \bar{\cV}^*$ and $V \simeq \bar{V}^*$.)
Note that from this condition and \eqref{eq:v-reality}, \eqref{eq:cr-psi} it follows that
\begin{equation} \label{eq:f-reality}
 f_{i+r}(\zeta) = \zeta \overline{f_i(-1/\bar\zeta)}.
\end{equation}

\end{itemize}

Given $\Psi(\zeta)$ obeying all the above conditions, all of the
holomorphic structures $V_\zeta$ indeed come from a single unitary connection $D$ in $V$.
To write $D$ explicitly let us work in a unitary trivialization of $V$.
From above, this trivialization determines functions $f_i^{(0)}$ and $f_i^{(1)}$.
We then write $D = \de + A$, with $A$ determined by
\begin{equation}
 \iota_{v_i^{(0)}} A = f_i^{(0)}, \qquad  \iota_{v_i^{(1)}} A = f_i^{(1)},
\end{equation}
or equivalently
\begin{equation} \label{eq:a-determined}
 \iota_{v_i(\zeta)} A = f_i(\zeta).
\end{equation}
The unitarity of $D$ follows from \eqref{eq:v-reality}, \eqref{eq:f-reality}.

\section{The semiflat case} \label{sec:semiflat}

So far we have constructed a complex unitary line bundle $V$ over $\cM$,
and explained our general strategy
for equipping it with a hyperholomorphic connection $D$.
In this section we begin implementing this strategy.

We first consider a simplified situation in which we neglect the instanton corrections coming from 4d BPS particles, i.e.
we take the \hk structure on $\cM$ to be the one arising from naive dimensional reduction
of the theory $T_4$.
The corresponding \hk metric was written down first in \cite{Cecotti:1989qn} and is sometimes called
the ``$c$-map metric''; we call it the ``semiflat metric''.
It is characterized \cite{Gaiotto:2008cd} by the condition that the functions
$\cY^\sf_\gamma$ given in \eqref{eq:y-sf} are complex Darboux coordinates.

\subsection{The hyperholomorphic connection}

In this case we can write down a hyperholomorphic connection in $V$ very simply:
\begin{align} \label{eq:a-sf}
 A^\sigma = A^{\sf,\sigma} = \frac{\I}{4 \pi} \IP{\theta^\sigma, \de \theta} + \frac{\I \pi R^2}{4} (\IP{Z, \de \bar{Z}} + \IP{\bar{Z}, \de Z}).
\end{align}
Note that this differs from \eqref{eq:A-sigma} only by the addition of the second term.  In particular,
it indeed descends to a connection in $V$, for the same reason that \eqref{eq:A-sigma} does.
The corresponding curvature is
\begin{equation} \label{eq:hh-curv-sf}
 F = F^\sf = \frac{\I}{4 \pi} \IP{\de \theta, \de \theta} + \frac{\I \pi R^2}{2} \IP{\de Z, \de \bar{Z}}.
\end{equation}
As befits the curvature of a hyperholomorphic connection,
$F$ is of type $(1,1)_\zeta$ for all $\zeta$; one could prove this directly, but we do not bother since
it is an automatic consequence of the construction we give in the next section.

$F$ differs from the \kahler form on $\cM$ in the complex structure
$J_{\zeta = 0}$ only by an irrelevant factor $2 \pi \I R$ and a crucial relative minus sign,
\begin{equation}
 2 \pi \I R \omega_{\zeta = 0} = - \frac{\I}{4 \pi} \IP{\de \theta, \de \theta} + \frac{\I \pi R^2}{2} \IP{\de Z, \de \bar{Z}}.
\end{equation}

\subsection{Twistorial construction} \label{sec:twistor-sf}

Now let us explain how the connection \eqref{eq:a-sf} can arise from the twistorial construction
we gave in Section \ref{sec:hyperhol-structures}.  The main ingredient in that construction
is the object $\Psi$ defined there.
In this case we can give $\Psi$ directly by an explicit formula (chosen teleologically to
produce the desired $A$ below):
\begin{equation} \label{eq:psi-sf}
 \Psi^\sigma = \Psi^{\sf,\sigma} = \exp \left[\frac{\I \pi R^2}{4} \left( \zeta^{-2} U + \zeta^2 \bar{U} \right) - \frac{R}{4} \left(\zeta^{-1} W^\sigma + \zeta \bar{W}^\sigma \right) \right],
\end{equation}
where
\begin{align}
U &= \int \IP{Z, \de Z}, \\
W^\sigma &= \IP{Z, \theta^\sigma}.
\end{align}
$U$ is a function on $\cB$ (through the upper limit
of the integral), which we think of as a function on $\cM$ by pullback.
It is insensitive to small deformations of the contour of integration thanks to \eqref{eq:lagrangian-condition},
but does depend
on a choice of base-point on $\cB$ (the lower limit of the integral) and in principle
on the homotopy class of the contour.  In all examples I know, $\pi_1(\cB)$ actually vanishes,
so there is no issue of homotopy; one does have to worry about the base-point, but this is just
a single choice made once and for all.
Having made this choice, $\Psi^\sigma$ becomes well
defined over $\cN^\sigma$, and descends to a globally defined section of $\Hom(\cV_\zeta, V)$ over $\cM$.\footnote{We
should check that \eqref{eq:psi-sf} indeed defines a section of $\Hom(\cV_\zeta, V)$.
We use the fact that the canonical logarithms $\xi^\sigma$ are given concretely by
the log of \eqref{eq:y-sf},
\begin{equation}
 \xi^\sigma = -\I \left[  \pi R (\zeta^{-1} Z + \zeta \bar{Z}) \right] + \theta^\sigma.
\end{equation}
Under the shift $T_\delta$, $\Psi^\sigma$ is transformed by a factor
\begin{equation}
\exp \left[- \frac{\pi R}{2} \left(\zeta^{-1} \IP{Z, \delta} + \zeta \IP{\bar{Z},\delta} \right) \right]
= \exp \left[ \frac{\I}{2} \IP{\theta^\sigma-\xi^\sigma, \delta} \right].
\end{equation}
This matches the expected transformation for a section of $\Hom(\cV_\zeta, V)$
according to \eqref{eq:gammastar-action}, \eqref{eq:gammastar-action-complex}.  Moreover, if $\sigma(\gamma) / \sigma'(\gamma) = (-1)^{c^{\sigma \sigma'} \cdot \gamma}$
then
\begin{equation}
 \Psi^{\sigma'} / \Psi^{\sigma} = \exp \left[ - \frac{\pi R}{4} \left( \zeta^{-1} \IP{Z, c^{\sigma \sigma'}} + \zeta \IP{\bar{Z}, c^{\sigma \sigma'}} \right) \right]
= \exp \left[ \frac{\I}{4} \IP{\theta^\sigma-\xi^\sigma, c^{\sigma \sigma'}} \right]
\end{equation}
which is also as it should be, according to
\eqref{eq:sig-shift-real-2}, \eqref{eq:sig-shift-complex-2}.}

The next important ingredient in the recipe of Section \ref{sec:hyperhol-structures} is a
basis for the $(0,1)$ vector fields on $\cM$.  For the particular $\cM$ we are now considering,
such a basis was given in \cite{Gaiotto:2008cd}:
\begin{align}
 v_i(\zeta) &= \I \pi R \frac{\partial Z}{\partial u^i} \cdot \frac{\partial}{\partial \theta} + \zeta \frac{\partial}{\partial u^i},  \quad i = 1, \dots, r, \\
v_{i+r}(\zeta) &= \frac{\partial}{\partial \bar{u}^{\bar{i}}} + \I \pi R \zeta \frac{\partial \bar{Z}}{\partial \bar{u}^{\bar{i}}} \cdot \frac{\partial}{\partial \theta}, \quad i = 1, \dots, r.
\end{align}
Applying these vector fields to $\Psi^\sigma$ we have
\begin{equation}
 (\bar\partial_{v_i(\zeta)} + f_i(\zeta)) \Psi = 0,
\end{equation}
where
\begin{align}
 f_i(\zeta) &= \frac{R}{4} \left\langle\frac{\partial Z}{\partial u^i}, \theta^\sigma\right\rangle + \frac{\I \pi R^2}{4} \zeta \left\langle\bar{Z}, \frac{\partial Z}{\partial u^i}\right\rangle, \quad i = 1, \dots, r, \label{eq:fi-sf-1} \\
 f_{i + r}(\zeta) &= \frac{\I \pi R^2}{4} \left\langle Z, \frac{\partial \bar Z}{\partial \bar{u}^{\bar{i}}}\right\rangle + \frac{R}{4} \zeta \left\langle\frac{\partial \bar Z}{\partial \bar u^{\bar i}}, \theta^\sigma\right\rangle, \quad i = 1, \dots, r. \label{eq:fi-sf-2}
\end{align}
Note that the $f_i$ are of the simple form \eqref{eq:fi-decomp}, so our twistorial construction of a hyperholomorphic connection
can be applied.  Moreover we compute directly that the connection \eqref{eq:a-sf} indeed obeys \eqref{eq:a-determined}
with this $f$.

So our twistorial construction starting from \eqref{eq:psi-sf}
has produced the desired hyperholomorphic connection $A$ in $V$.

\section{$U(1)$ theory with $1$ hypermultiplet} \label{sec:ov}

So far we have applied our twistorial construction of a hyperholomorphic connection $D$
to the simple ``semiflat'' case.
Now we begin to consider the inclusion of quantum corrections coming from BPS particles of the 4d theory:  in other words we confront the case where the \hk structure on $\cM$ is determined by
the exact equation \eqref{eq:X-integral-mult-explicit} rather than the approximation
\eqref{eq:y-sf}.

As in \cite{Gaiotto:2008cd}, the only cases which we can study exactly are ones in which the BPS particles carry only electric charges (in some duality frame),
as opposed to the usual situation where we have both electrically
and magnetically charged BPS particles (in every duality frame.)

The most fundamental example we can study exactly arises by taking $T_4$
to be the $\N=2$ supersymmetric Yang-Mills theory with gauge group $U(1)$,
coupled to a single matter hypermultiplet of charge $1$.  This theory is not asymptotically free,
but that does not pose a real obstacle for the analysis we are doing here (it does mean that the metric on
$\cM$ will be incomplete.)
The \hk metric on $\cM$ in this case (sometimes called ``Ooguri-Vafa metric'' or ``periodic NUT space'')
was described in \cite{Ooguri:1996me,Seiberg:1996ns}, and reconsidered in
\cite{Gaiotto:2008cd} as an example of the general twistorial construction of \hk metrics there.
In this section
we extend that discussion to include the hyperholomorphic line bundle $V$ over $\cM$.

We will first write down an explicit formula for a hyperholomorphic connection $D$, and
then verify that our twistorial construction reproduces that connection.  One important
consequence of this result is that the connection $D$ produced by the twistorial construction
actually extends even over the naively ``singular'' locus where the torus fibers of $\cM$
degenerate, although $D^\sf$ does not.
This is similar to what happened in the construction of $\cM$ itself in
\cite{Gaiotto:2008cd}:  the BPS instanton corrections smooth out the singularities.

\subsection{Review}

First we recall the data of the theory $T_4$
and the structure of $\cM$.  The Coulomb branch $\cB$ is
the open disc
\begin{equation}
 \cB = \{a: \abs{a} < \abs{\Lambda} \} \subset \C.
\end{equation}
The discriminant locus $\cB_\sing$ consists of the single point $a=0$; at this point
the effective $U(1)$ theory with the hypermultiplet integrated out becomes singular.
The charge lattice $\Gamma$ has rank 2, with
local generators $\gamma_\e$ and $\gamma_\m$ obeying $\IP{\gamma_\m,\gamma_\e} = 1$,
and counterclockwise monodromy around $a=0$ given by
\begin{equation} \label{eq:ov-monodromy}
 \gamma_\e \mapsto \gamma_\e, \quad
 \gamma_\m \mapsto \gamma_\m + \gamma_\e.
\end{equation}
(So $\gamma_\e$ is a global section of $\Gamma$, but $\gamma_\m$ only exists locally.)  The central charges are
\begin{equation} \label{eq:Z-ov}
 Z_{\gamma_\e} = a, \quad Z_{\gamma_\m} = \frac{1}{2 \pi \I} (a \log \frac{a}{\Lambda} - a).
\end{equation}

Our discussion in Section \ref{sec:compactification} says $\cM$ is fibered over the disc $\cB$, with the generic fiber a 2-torus
coordinatized by $\varphi_\e$ and $\varphi_\m$, and with some kind of degenerate fiber over $a=0$.
Since $\gamma_\e$ is monodromy invariant, the coordinate $\varphi_\e$ actually exists globally,
i.e. we have a global circle factor in $\cM$:  so we can
view $\cM$ as fibered over the three-dimensional base $\cB \times S^1$, with the generic fiber a circle,
but with potential degenerate fibers over the circle $\{a=0\} \times S^1$.

To write the \hk metric on $\cM$ explicitly, it is convenient to construct it on a covering space $\cN$
of $\cM$, in such a way that it descends to $\cM$.
More precisely, we already described in Section \ref{sec:bundle-def} a canonical covering space $\cN^\sigma$
associated to a choice of quadratic refinement $\sigma$.
So, let $\sigma$ be the quadratic refinement determined by
\begin{equation}
 \sigma(\gamma_\e) = -1, \qquad \sigma(\gamma_\m) = 1.
\end{equation}
This $\sigma$ is actually invariant under the monodromy \eqref{eq:ov-monodromy} around $a=0$,
hence is globally defined over $\cB$, which simplifies our life:  we never have to consider any other quadratic refinement,
and so in the rest of Section \ref{sec:ov} we can drop the superscript $\sigma$.
The covering space $\cN$ has coordinates $a \in \C$, $\theta_\e \in \R$ and $\theta_\m \in \R$;
$a$ and $\theta_\e$ are global, while $\theta_\m$ is only local (as it depends on a choice of $\gamma_\m$).
We will view $\cN$ as an affine $\R$-bundle over the base $\cB \times \R$
coordinatized by $a$, $\theta_\e$.

Since the 4d theory has no magnetically charged BPS particles, there are no instanton corrections that depend on $\theta_\m$,
and hence shifts of $\theta_\m$ are isometries of $\cN$.  This implies that
the \hk metric in $\cN$ is of the ``Gibbons-Hawking'' form:
it is determined by a positive harmonic function $H$ on $\cB \times \R$ (with singularities)
and a connection in $\cN$ with connection 1-form $\Theta$.
The Gibbons-Hawking form of the metric is
\begin{equation} \label{eq:g-h}
g = H \de \vec{x}^2 + H^{-1} \Theta^2, \qquad
\end{equation}
where we introduced a vector $\vec{x}$ on the base $\cB \times \R$ by
\begin{equation}
a = x^1 + \I x^2, \quad \theta_\e = 2 \pi R x^3.
\end{equation}
Any metric of the form \eqref{eq:g-h} is known to be \hk, provided that the curvature of the connection obeys
\begin{equation}
\de \Theta = \star \de H.
\end{equation}
In order for the metric to descend to the quotient $\cM$,
we require further
\begin{equation}
 H(a, \theta_\e + 2 \pi) = H(a, \theta_\e), \qquad B(a, \theta_\e + 2 \pi) = B(a, \theta_\e).
\end{equation}

The explicit function $H$ in our case is
\begin{equation} \label{eq:V-sum}
H = \frac{R}{4 \pi} \sum_{n=-\infty}^\infty \left( \frac{1}{\sqrt{R^2 \abs{a}^2 + (\frac{\theta_\e}{2 \pi} + n + \half)^2}} - \kappa_n \right).
\end{equation}
Here the $\kappa_n$ are some irrelevant constants, whose only purpose is to ensure
convergence of the sum.
In particular, \eqref{eq:V-sum} says $H$ is singular only at the periodic array of points $a = 0$, $\theta_\e = 2 \pi (n + \half)$.
Poisson resummation gives $H = H^\sf + H^\inst$, where (for appropriate $\kappa_n$)
\begin{align}
H^\sf &= -\frac{R}{4\pi} \left(\log \frac{a}{\Lambda} + \log \frac{\bar{a}}{\bar{\Lambda}} \right), \\
H^\inst &= \frac{R}{2\pi} \sum_{n\neq 0} e^{\I n \theta_\e} K_0(2 \pi R \abs{n a}). \label{eq:Hinst}
\end{align}
Here and below, $K_k$ denotes the $k$-th modified Bessel function.

Locally on $\cB$, when we make a choice of generator $\gamma_\m$
as above, we get a local coordinate $\theta_\m$ which trivializes the affine $\R$-bundle $\cN$.
Then we can represent the connection as
\begin{equation} \label{eq:Theta}
 \Theta = \frac{\de \theta_\m}{2 \pi} + B,
\end{equation}
with $B$ a locally defined $1$-form on $\cB \times \R$.  This will be convenient for
writing concrete formulas.  In our case $B$
decomposes as $B = B^\sf + B^\inst$ with
\begin{align}
B^\sf &= \frac{\I}{8 \pi^2} \left( \log \frac{a}{\Lambda} - \log \frac{\bar{a}}{\bar{\Lambda}} \right) \de \theta_\e, \label{eq:b-sf} \\
B^\inst &=  - \frac{R}{4 \pi} \left( \frac{\de a}{a} - \frac{\de \bar a}{\bar a} \right) \sum_{n \neq 0} e^{\I n \theta_\e} \abs{a} (-1)^n \sgn(n) K_1(2 \pi R \abs{n a}).  \label{eq:Binst}
\end{align}
(The need to choose a branch of the logarithm
in \eqref{eq:b-sf} reflects the fact that $B$ is only locally defined.  Indeed, changing the branch shifts
$B$ by $k \de \theta_\e / 2 \pi$, for some $k \in \Z$.  This shift corresponds
to changing our choice of $\gamma_\m$ by $\gamma_\m \to \gamma_\m - k \gamma_\e$, giving
$\theta_\m \to \theta_\m - k \theta_\e$, so that $\Theta$ in
\eqref{eq:Theta} is invariant and globally defined.)

\subsection{The hyperholomorphic bundle $V$} \label{sec:hh-direct-ov}

Now we describe a hyperholomorphic line bundle $V$ over the Ooguri-Vafa space $\cM$.

As a Hermitian line bundle, $V$ is constructed according to our usual recipe from Section \ref{sec:complex-line-bundle}.
So $V$ is the quotient of $\cN \times \C$ by the $\Gamma^*$ action with generators
\begin{equation}
 T_\e (\theta_\e, \theta_\m, \psi) = (\theta_\e + 2 \pi, \theta_\m, \psi e^{\frac{\I}{2} \theta_\m}), \quad T_\m (\theta_\e, \theta_\m, \psi) = (\theta_\e, \theta_\m  + 2 \pi, - \psi e^{-\frac{\I}{2} \theta_\e}).
\end{equation}
Our hyperholomorphic connection in $V$ is of the form
\begin{equation} \label{eq:a-ov}
 A = - \I \frac{P}{H} \Theta + \I \alpha + \frac{\I}{4\pi} \de (\theta_\e \theta_\m)
\end{equation}
where $P$ is a harmonic function on $\cB \times \R$,
and $\alpha$ is a 1-form on $\cB \times \R$ obeying
\begin{equation}
 \de \alpha = \star \de P
\end{equation}
(such 1-forms exist since $P$ is harmonic).
It is a straightforward exercise to show that if $\cN$ is any Gibbons-Hawking space,
i.e. a space with metric \eqref{eq:g-h},
with the 1-form $A$ given by \eqref{eq:a-ov}, then $F = \de A$ is indeed of type $(1,1)_\zeta$ for all $\zeta \in \C\PP^1$.
In our case $P$ will be engineered moreover to obey
\begin{equation}
 P(a, \theta_\e + 2 \pi) = P(a, \theta_\e) + 2 \pi H(a, \theta_\e),
\end{equation}
which guarantees that $A$ indeed descends to a connection in $V$ over $\cM$.

The desired $P$ can be given explicitly in the form
\begin{equation}
 P(a, \theta_\e) = - \frac{R}{2} \sum_{n=-\infty}^\infty \left( \frac{n + \half}{\sqrt{R^2 \abs{a}^2 + (\frac{\theta_\e}{2 \pi} + n + \half)^2}} - \kappa'_n \theta_\e \right),
\end{equation}
which Poisson resums into $P = P^\sf + P^\inst$, with
\begin{align}
P^\sf &= -\frac{R \theta_\e}{4\pi} \left(\log \frac{a}{\Lambda} + \log \frac{\bar{a}}{\bar{\Lambda}} \right) = \theta_\e H^\sf, \label{eq:p-sf} \\
P^\inst &= \frac{R}{2\pi} \sum_{n \neq 0} e^{\I n \theta_\e} (-1)^n \left[ \theta_\e K_0(2 \pi R \abs{na}) + 2 \pi \I R \abs{a} \sgn(n) K_1(2 \pi R \abs{na}) \right]. \label{ginst}
\end{align}

The corresponding $\alpha$ is also of the form $\alpha = \alpha^\sf + \alpha^\inst$, where
\begin{multline}
\alpha^\sf = \frac{\I \theta_\e}{8 \pi^2} \left( \log \frac{a}{\Lambda} + \log \frac{\bar{a}}{\bar{\Lambda}} \right) \de \theta_\e \\- \frac{\I R^2}{8} \left( (a \log \frac{a}{\Lambda} + a \log \frac{\bar{a}}{\bar\Lambda} - a) \de \bar{a} - (\bar{a} \log \frac{\bar{a}}{\bar{\Lambda}} - \bar{a} \log \frac{a}{\Lambda} - \bar{a}) \de a \right), \label{eq:alpha-sf}
\end{multline}
but we do not write $\alpha^\inst$ explicitly.

By an appropriate change of coordinates one can check directly that the hyperholomorphic
connection actually extends smoothly even over the naively ``singular'' points where
$a=0$ and $\theta_\e = 2\pi(n + \half)$.

\subsection{Twistorial construction} \label{sec:twistor-ov}

Now we explain how our general twistorial construction can be applied to produce
the exact hyperholomorphic connection $A$.  Following the discussion
of Section \ref{sec:identifying}, our main job is to give an appropriate ``instanton-corrected'' version of the map $\Psi$.
To write the desired $\Psi$ we first recall the instanton-corrected $\cY$ given in \cite{Gaiotto:2008cd}.
We use the notation $\xi_\e$, $\xi_\m$ for $\xi_{\gamma_\e}$, $\xi_{\gamma_\m}$ respectively.
Then
\begin{align}
 \xi_\e &= \xi^\sf_\e, \\
 \xi_\m &= \xi^\sf_\m + \xi^\inst_\m,
\end{align}
where
\begin{align}
\xi^\sf_\e &= -\I \pi R \zeta^{-1} a + \theta_\e -\I \pi R \zeta \bar{a} , \label{eq:ysfe-ov} \\
\xi^\sf_\m &= - \frac{1}{2} R \zeta^{-1} (a \log \frac{a}{\Lambda} - a) + \theta_\m + \frac{1}{2} R \zeta (\bar a \log \frac{\bar a}{\bar \Lambda} - \bar a) , \\
\xi^\inst_\m &= \frac{1}{4\pi} \int_{\ell_+} \frac{\de \zeta'}{\zeta'}
\frac{\zeta' + \zeta}{\zeta' - \zeta} \log[1-\cY_\e(\zeta')]
- \frac{1}{4\pi} \int_{\ell_-}
 \frac{\de \zeta'}{\zeta'} \frac{\zeta' + \zeta}{\zeta' - \zeta} \log[1-\cY_{-\e}(\zeta')] , \label{eq:yminst-ov}
\end{align}
with the contours of integration $\ell_\pm$ defined by
\begin{equation}
 \ell_\pm = \pm a \R_- \subset \C.
\end{equation}
The integrals over $\ell_\pm$ reflect the instanton corrections due to 4d BPS particles of charge $\pm 1$.

We claim that the right formula for the corrected $\Psi$ is
\begin{equation}
 \Psi = \Psi^\sf \Psi^\inst
\end{equation}
where $\Psi^\sf$ was given in \eqref{eq:psi-sf} and
\begin{equation} \label{eq:psiinst-ov}
\begin{split}
\Psi^\inst(\zeta) = \exp\Biggl[
&- \frac{1}{16\pi^2} \int_{\ell_+} \frac{\de \zeta'}{\zeta'} \frac{\zeta' + \zeta}{\zeta' - \zeta} \left[ 2 \Li_2(\cY_\e(\zeta')) + (2 \I \xi_\e(\zeta') - \I \xi_\e(\zeta)) \log (1 - \cY_\e(\zeta')) \right] \\
&- \frac{1}{16\pi^2} \int_{\ell_-} \frac{\de \zeta'}{\zeta'} \frac{\zeta' + \zeta}{\zeta' - \zeta} \left[ 2 \Li_2(\cY_{-\e}(\zeta')) + (2 \I \xi_{-\e}(\zeta') - \I \xi_{-\e}(\zeta)) \log (1 - \cY_{-\e}(\zeta')) \right]
\Biggr].
\end{split}
\end{equation}
Here we take the \ti{principal branch} of both $\log$ and $\Li_2$,
using the fact that $\abs{\cY_{\pm \e}} < 1$ along the rays $\ell_\pm$.

First we should verify that $\Psi$ so defined is a section of $\Hom(\cV_\zeta, V) = \Hom(\cY(\zeta)^* \cV, V)$.  Begin by considering $\zeta$ away from the critical
rays $\ell_{\pm}$.
Recalling that $\Psi^\sf$ is a section of $\Hom(\cY^\sf(\zeta)^* \cV, V)$, we need to check
that $\Psi^\inst$ is a section of $\Hom(\cY(\zeta)^* \cV, \cY^\sf(\zeta)^* \cV)$:  or, in more human-readable terms, we need to check
\begin{align}
 \Psi^\inst(\theta_\e + 2\pi, \theta_\m) &= \Psi^\inst(\theta_\e, \theta_\m) \exp \left(-\frac{\I}{2} \xi^\inst_\m\right), \label{eq:psiinst-shift1-ov} \\
 \Psi^\inst(\theta_\e, \theta_\m + 2\pi) &= \Psi^\inst(\theta_\e, \theta_\m). \label{eq:psiinst-shift2-ov}
\end{align}
Indeed, under $\theta_\e \to \theta_\e + 2 \pi$ the terms $\xi_{\pm\e}(\zeta')$ and $\xi_{\pm\e}(\zeta)$
in \eqref{eq:psiinst-ov} are both shifted by $\pm 2 \pi$, while all other terms are unchanged.
This gives an explicit integral formula for the shift of $\Psi^\inst$, and using
\eqref{eq:yminst-ov} we see that this shift is the desired
\eqref{eq:psiinst-shift1-ov}.  \eqref{eq:psiinst-shift2-ov} is automatic since
$\Psi^\inst$ has no dependence on $\theta_\m$ at all.
In addition we need to verify that $\Psi$ indeed jumps by the canonical dilogarithm
$L_{\pm \e}$ at the rays $\ell_{\pm}$; but this follows directly from \eqref{eq:psiinst-ov}
by computing the residue.

Now we consider the construction of $A$ given in Section \ref{sec:identifying}.
One basic ingredient in the construction is a nice basis for $T^{0,1}\cM$; an explicit such
basis was given in \cite{Gaiotto:2008cd}:
\begin{align}
v_1(\zeta) &= (- \I \pi R \partial_{\theta_\e} + \pi (H + 2 \pi \I R B_{\theta_\e}) \partial_{\theta_\m}) + \zeta (2 \pi B_a \partial_{\theta_\m} - \partial_a), \label{eq:v1} \\
v_2(\zeta) &= (2 \pi B_{\bar a} \partial_{\theta_\m} - \partial_{\bar a}) + \zeta (-\I \pi R \partial_{\theta_\e} - \pi (H - 2 \pi \I R B_{\theta_\e}) \partial_{\theta_\m}). \label{eq:v2}
\end{align}
Note that they obey the reality condition \eqref{eq:v-reality} which in this case says
\begin{equation} \label{eq:v-rel-ov}
 v_2(\zeta) = \zeta \overline{v_1(-1/\bar\zeta)}.
\end{equation}
This being so, we focus just on $v_1$.
Breaking everything into its semiflat and instanton parts, we can write
\begin{equation}
 (\bar\partial_{v_1(\zeta)} \Psi + f_1^\sf(\zeta) + f_1^\inst(\zeta)) \Psi = 0
\end{equation}
where
\begin{align}
 f_1^\sf(\zeta) &= (- \I \pi R \partial_{\theta_\e} + \pi (H^\sf + 2 \pi \I R B^\sf_{\theta_\e}) \partial_{\theta_\m}) \log \Psi^\sf + \zeta (- \partial_a) \log \Psi^\sf,\\
 f_1^\inst(\zeta) &= (-\I \pi R \partial_{\theta_\e} - \zeta \partial_a) \log \Psi^\inst + (\pi H^\inst + 2 \pi \zeta B_a^\inst) \partial_{\theta_\m} \log \Psi^\sf. \label{eq:f1inst}
\end{align}
Let us write the instanton-correction term $f_1^\inst$
more explicitly.  For the first term in \eqref{eq:f1inst} we use the explicit form of $\Psi^\inst$ from \eqref{eq:psiinst-ov},
and the fact that
\begin{equation}
 (-\I \pi R \partial_{\theta_\e} - \zeta \partial_a) \cY_\e(\zeta') = \pi R (1 - \zeta/\zeta') \cY_\e(\zeta').
\end{equation}
For the second term we use \eqref{eq:psi-sf} to see that
\begin{equation}
 \partial_{\theta_\m} \log \Psi^\sf = \frac{R}{4} (\zeta^{-1} a + \zeta \bar{a}).
\end{equation}
Altogether \eqref{eq:f1inst} becomes
\begin{align}
 f_1^\inst(\zeta) = \frac{R}{16\pi} & \int_{\ell_+} \frac{\de \zeta'}{\zeta'} (1 + \zeta/\zeta') (\xi_\e(\zeta) - 2 \xi_\e(\zeta')) \frac {\cY_\e(\zeta')}{1 - \cY_\e(\zeta')} \\
- \frac{R}{16\pi} & \int_{\ell_-} \frac{\de \zeta'}{\zeta'} (1 + \zeta/\zeta') (\xi_{-\e}(\zeta) - 2 \xi_{-\e}(\zeta')) \frac {\cY_{-\e}(\zeta')}{1 - \cY_{-\e}(\zeta')} \\
 + \frac{R}{4} & (\pi H^\inst + 2 \pi \zeta B^\inst_a) (\zeta^{-1} a + \zeta \bar{a}).
\end{align}
Expanding the geometric series, writing out $\xi_\e$ explicitly using \eqref{eq:ysfe-ov},
and inserting $H^\inst$ and $B^\inst$ from \eqref{eq:Hinst} and \eqref{eq:Binst}, this becomes
\begin{align}
 f_1^\inst(\zeta) = \frac{R}{16\pi} & \int_{\ell_+} \frac{\de \zeta'}{\zeta'} (1 + \zeta/\zeta') (\pi R (\zeta^{-1}a - 2 \zeta'^{-1} a + \zeta \bar a - 2 \zeta' \bar a) - \I \theta_\e) \sum_{n>0} \cY_\e(\zeta')^n \\
 + \frac{R}{16\pi} & \int_{\ell_-} \frac{\de \zeta'}{\zeta'} (1 + \zeta/\zeta') (\pi R (\zeta^{-1}a - 2 \zeta'^{-1} a + \zeta \bar a - 2 \zeta' \bar a) - \I \theta_\e) \sum_{n>0} \cY_{-\e}(\zeta')^n  \\
 + \frac{R^2}{8} & (\zeta^{-1} a + \zeta \bar{a}) \sum_{n\neq 0} e^{\I n \theta_\e} \left( \sgn(n) K_0(2 \pi R \abs{n a}) - \zeta \frac{\abs{a}}{a} K_1(2 \pi R \abs{n a}) \right).
\end{align}
We evaluate the integrals using
\begin{align}
 \int_{\ell_+} \frac{\de \zeta'}{\zeta'} (\zeta')^k \cY_\e(\zeta')^n &= 2 (-1)^n \left( - \frac{a}{\abs{a}} \right)^k K_k(2 \pi R n \abs{a}) e^{\I n \theta_\e}, \\
 \int_{\ell_-} \frac{\de \zeta'}{\zeta'} (\zeta')^k \cY_{-\e}(\zeta')^n &= 2 (-1)^n \left( \frac{a}{\abs{a}} \right)^k K_k(2 \pi R n \abs{a}) e^{-\I n \theta_\e},
\end{align}
and collect terms to get
\begin{equation} \label{eq:f1inst-final}
 f_1^\inst(\zeta) = \sum_{n \neq 0} e^{\I n \theta_\e} \left[ \frac{R^2}{8} \left( 4 \abs{a} K_1 - 2 \zeta \bar{a} (K_0 + K_2) \right) - \frac{\I R}{8 \pi} \theta_\e \left(K_0 - \zeta \frac{\abs{a}}{a} K_1\right) \right],
\end{equation}
where we have adopted the shorthand
\begin{equation}
 K_k = (\sgn(n))^k (-1)^n K_k(2 \pi R \abs{n a}).
\end{equation}
$f_2(\zeta)$ can be determined from $f_1(\zeta)$ using the relation \eqref{eq:v-rel-ov}.

Finally, now that we have determined the two functions $f_i(\zeta)$, we could
use the recipe of Section \ref{sec:hyperhol-structures}
to construct the hyperholomorphic connection $A$.
(For this it is important that the terms proportional to $\zeta^{-1}$ and $\zeta^2$ which appeared at intermediate steps have cancelled
in the final result \eqref{eq:f1inst-final}.)
We do not write the resulting $A$ explicitly here, but in the next section we will
determine it.

\subsection{Comparing the constructions}

In the last two subsections we have described the hyperholomorphic connection $A$ in two different ways.
In Section \ref{sec:hh-direct-ov} we wrote a formula for the connection directly; in Section \ref{sec:twistor-ov} we gave a
twistorial recipe for the connection but stopped short of using it to produce an explicit formula.
In this section we check that these two descriptions indeed describe the \ti{same} connection.

We first consider the semiflat approximations to the two descriptions.
In this approximation the connection \eqref{eq:a-ov} simplifies:  we have $P = P^\sf$, $B = B^\sf$,
$\alpha = \alpha^\sf$,
and a direct computation from \eqref{eq:a-ov} using \eqref{eq:p-sf}, \eqref{eq:b-sf}, \eqref{eq:a-sf}
then gives
\begin{equation} \label{eq:a-sf-direct}
 A = \frac{\I}{4 \pi} (\theta_\m \de \theta_\e - \theta_\e \de \theta_\m) + \frac{R^2}{8} \left( (a \log \frac{a}{\Lambda} + a \log \frac{\bar a}{\bar \Lambda} - a) \de \bar{a} - (\bar{a} \log \frac{\bar{a}}{\bar{\Lambda}} + \bar{a} \log \frac{a}{\Lambda} - \bar{a}) \de a \right).
\end{equation}
On the other hand, in Section \ref{sec:twistor-sf} we considered the semiflat approximation
of our general twistorial construction,
and showed that in this limit the connection $A$ is given by \eqref{eq:a-sf}.  Using \eqref{eq:Z-ov} we see that
\eqref{eq:a-sf} indeed agrees with \eqref{eq:a-sf-direct}.  So the two descriptions agree in the semiflat approximation.

Now let us compare the two descriptions of the instanton corrections to $A$.
More precisely, we will check that the corrections to one component of $A$ match, namely
$\iota_{\partial_{\theta_\m}} A$.  This is the simplest thing to look at, in particular because
it does not involve the 1-form $\alpha$ which we have not written explicitly.

It turns out that the formulas will be shorter if, rather than writing the correction to $\iota_{\partial_{\theta_\m}} A$ directly, we
look at the correction to $\pi H \iota_{\partial_{\theta_\m}} A$.
Using our explicit description \eqref{eq:a-ov} of $A$
we can read off this correction as:
\begin{equation}
 \pi (H \iota_{\partial_{\theta_\m}} A)^\inst = -\frac{\I}{2} P^\inst + \frac{\I}{4} \theta_\e H^\inst = \sum_{n \neq 0} e^{\I n \theta_\e} \left[ \frac{R^2}{2} \abs{a} K_1 -  \frac{\I R}{8 \pi} \theta_\e K_0 \right].
\end{equation}
On the other hand, from our twistorial construction and
\eqref{eq:v1}, we see that $\pi H \iota_{\partial_{\theta_\m}} A$ is the real part of the $\zeta$-independent part of $f_1(\zeta)$.
Then \eqref{eq:f1inst-final} determines the instanton correction
to this quantity as
\begin{equation}
 \pi (H \iota_{\partial_{\theta_\m}} A)^\inst = \sum_{n \neq 0} e^{\I n \theta_\e} \left[ \frac{R^2}{2} \abs{a} K_1 - \frac{\I R}{8 \pi} \theta_\e K_0 \right].
\end{equation}
The two agree perfectly.

\section{The general case} \label{sec:general}

In the last two sections we have considered simple cases of our general construction,
where everything (both the
\hk metric in $\cM$ and the hyperholomorphic connection $A$ in $V$) can be computed explicitly.
Now let us return to the more general setting of Section \ref{sec:review}.
In that case, as already noted in \cite{Gaiotto:2008cd}, the story is not quite so computable.
In particular, the crucial functions $\cY_\gamma$ can only be determined implicitly as solutions of the
integral equation \eqref{eq:X-integral-mult-explicit}.
Nevertheless, as we now show, one can still construct a hyperholomorphic connection in $V$.

We take $\Psi$ to be
\begin{equation} \label{eq:psitotal}
 \Psi = \Psi^\sf \Psi^\inst,
\end{equation}
where $\Psi^\sf$ was given in \eqref{eq:psi-sf}, and
\begin{equation} \label{eq:psiinst-pieces}
 \Psi^{\inst,\sigma} = \Psi^{\inst,\sigma,1} \Psi^{\inst,\sigma,2},
\end{equation}
with
\begin{align} \label{eq:psiinst1}
 \Psi^{\inst,\sigma,1} &= \exp \left[ - \sum_\gamma \frac{\Omega(\gamma)}{16 \pi^2} \int_{\ell_\gamma} \frac{\de \zeta'}{\zeta'} \frac{\zeta'+\zeta}{\zeta'-\zeta}
\left[ 2 \Li_2(\cY_\gamma(\zeta')) + \I \xi^\sigma_\gamma(\zeta') \log (1 - \cY_\gamma(\zeta')) \right]\right],\\ \label{eq:psiinst2}
 \Psi^{\inst,\sigma,2} &= \exp \left[ - \sum_\gamma \frac{\Omega(\gamma)}{16 \pi^2} \int_{\ell_\gamma} \frac{\de \zeta'}{\zeta'} \frac{\zeta'+\zeta}{\zeta'-\zeta}
\left[ \I (\xi^{\sigma,\sf}_\gamma(\zeta') - \xi^{\sigma,\sf}_\gamma(\zeta)) \log (1 - \cY_\gamma(\zeta')) \right]\right].
\end{align}

The equations \eqref{eq:psiinst-pieces}-\eqref{eq:psiinst2} generalize \eqref{eq:psiinst-ov},
albeit in a rather peculiar-looking way.  The two parts of $\Psi^\inst$ play rather different roles.
The piece $\Psi^{\inst,1}$ is responsible for correcting $\Psi$ from a section of
$\Hom(\cY^\sf(\zeta)^* \cV, V)$ to a section of
$\Hom(\cY(\zeta)^* \cV, V)$.
The piece $\Psi^{\inst,2}$ is responsible for adjusting the \ti{asymptotics}
of $\Psi$ as $\zeta \to 0$ or $\zeta \to \infty$.

We claim that using this $\Psi$ in the twistorial construction of Section
\ref{sec:hyperhol-structures} leads to a hyperholomorphic connection $D$ in $V$.
Unlike the two cases we considered so far,
here we will not check this claim by directly computing $D$.
Instead we will just show that $\Psi$ obeys the conditions
of our twistorial construction, set out in Section \ref{sec:hyperhol-structures},
and in particular check that $\Psi$ does not jump at the walls of marginal stability in $\cB$.

\subsection{Asymptotics}

Recall from Section \ref{sec:identifying} that we define functions $f_i(\zeta)$ on $\cM$ by
\begin{equation} \label{eq:fdef}
\left( \bar\partial_{v_i(\zeta)} + f_i(\zeta) \right) \Psi(\zeta) = 0 \qquad (i = 1, \dots, 2 r).
\end{equation}
These functions determine the holomorphic structure in $V_\zeta$ which we are after.
As explained in Section \ref{sec:identifying}, in order to see that the holomorphic structures
in $V_\zeta$ indeed come from a hyperholomorphic connection,
what we need to show is that the $\zeta$ dependence of $f_i(\zeta)$ is in the simple form
\begin{equation} \label{eq:fi-simple}
f_i(\zeta) =  f^{(0)}_i +  \zeta f^{(1)}_i.
\end{equation}
We have already computed directly that the functions $f_i^\sf$
defined by
\begin{equation} \label{eq:fisf}
\left( \bar\partial_{v^\sf_i(\zeta)} + f^\sf_i(\zeta) \right) \Psi^\sf(\zeta) = 0 \qquad (i = 1, \dots, 2 r)
\end{equation}
are indeed of this form, as shown in \eqref{eq:fi-sf-1}, \eqref{eq:fi-sf-2}.
Our aim is to transfer this to a statement about the corrected $f_i$.

For this purpose it is convenient to introduce an auxiliary object
$\Upsilon = (\cY^\sf)^{-1} \circ \cY$.  This is a map $\cM \to \cM_\C$ which encapsulates
all of the quantum corrections to the \hk structure of $\cM$.
One of its key properties, exploited heavily in \cite{Gaiotto:2008cd}, is that $\Upsilon$
has a finite limit $\Upsilon_0$ as $\zeta \to 0$, and similarly $\Upsilon_\infty$
as $\zeta \to \infty$.  This fact gives us some control
over the effect of the quantum corrections.

Formally, pulling back the whole equation \eqref{eq:fisf} by $\Upsilon$ gives
\begin{equation} \label{eq:fisf-2}
\left( \bar\partial_{\Upsilon^* v_i^\sf(\zeta)} + \Upsilon^* f^\sf_i(\zeta) \right) \Upsilon^* \Psi^\sf(\zeta) = 0 \qquad (i = 1, \dots, 2 r).
\end{equation}
This is not yet the desired equation \eqref{eq:fdef}.  But let us consider its $\zeta \to 0$
limit.  We do have $\Upsilon_0^* ((v_i^\sf)^{(0)}) = v^{(0)}_i$ \cite{Gaiotto:2008cd}.
If we had $\Psi = \Upsilon^*\Psi^\sf$, we would conclude immediately
that $f_i(\zeta)$ is finite in the $\zeta \to 0$ limit (and indeed
that $f_i^{(0)} = \Upsilon_0^*(f_i^\sf)^{(0)}$.)
The actual relation between $\Psi$ and $\Upsilon^* \Psi^\sf$ is a bit subtler:  as we will
show below, they are
not equal, but they differ only by a finite correction as $\zeta \to 0$, i.e.
\begin{equation} \label{eq:good-asymptotics}
 \lim_{\zeta \to 0} \frac{\Psi}{\Upsilon^*\Psi^\sf} \text{ exists.}
\end{equation}
This is enough to show that $f_i(\zeta)$ is of the desired form \eqref{eq:fi-simple}.

Now, why is \eqref{eq:good-asymptotics} true?
We begin from the formula \eqref{eq:psitotal} for $\Psi$.
The term $\Psi^{\inst,1}$ is finite in the $\zeta \to 0$
limit, but $\Psi^{\inst,2}$ is not; indeed, taking $\zeta \to 0$
in \eqref{eq:psiinst2} gives
\begin{equation}
 \Psi^{\inst,\sigma,2} \sim c \exp \left[ \frac{1}{\zeta} \frac{\I R}{16 \pi} \sum_\gamma Z_\gamma  \Omega(\gamma) \int_{\ell_\gamma} \frac{\de \zeta'}{\zeta'}
\log (1 - \cY_\gamma(\zeta')) \right].
\end{equation}
On the other hand, using the definition \eqref{eq:psi-sf} of $\Psi^\sf$, we also know that
as $\zeta \to 0$ we have
\begin{align}
 \Upsilon^* \Psi^{\sf,\sigma} / \Psi^{\sf,\sigma} &\sim c \exp \left[-\frac{1}{\zeta} \frac{R}{4} \IP{Z,\Upsilon_0^* \theta - \theta} \right] \\
& = c \exp \left[\frac{1}{\zeta} \frac{\I R}{16 \pi} \sum_\gamma Z_\gamma \Omega(\gamma)
\int_{\ell_\gamma} \frac{\de \zeta'}{\zeta'}
\log (1 - \cY_\gamma(\zeta'))  \right].
\end{align}
Combining these results we get the desired \eqref{eq:good-asymptotics}.

\subsection{Wall-crossing} \label{sec:wall-crossing}

Now we consider the discontinuity properties of $\Psi$.

First we hold $x \in \cM$ fixed and consider $\Psi(x, \zeta)$ as a function of $\zeta$.
Given any ray $\ell$ in the complex plane we can consider the limits $\Psi^+$, $\Psi^-$ of $\Psi$
as $\zeta$ approaches $\ell$ clockwise, counterclockwise respectively.
From \eqref{eq:psiinst-pieces}-\eqref{eq:psiinst2} it follows that the two limits are related by
\begin{equation}
 \Psi^{+,\sigma} = \Psi^{-,\sigma} \prod_{\gamma: \ell_\gamma = \ell} \exp \left[ \frac{\I}{4 \pi} \Omega(\gamma) \left[ 2 \Li_2(\cY_\gamma) + \I \xi^\sigma_\gamma \log (1 - \cY_\gamma) \right]\right].
\end{equation}
So there are some special rays $\ell$ where $\Psi^\sigma$ jumps.
This is not unexpected ---
in fact, as we have described,
these are exactly the jumps required if $\Psi$ is to be a section of $\Hom(\cV_\zeta, V)$.

Now what about holding $\zeta \in \C^\times$ fixed and varying $x$?  For a generic small variation of $x$,
all of the ingredients in \eqref{eq:psiinst-pieces}-\eqref{eq:psiinst2} vary continuously, hence so does
$\Psi^\sigma$.  However, there is one point that needs attention:  on $\cB$ there are
codimension-1 loci at which several of the rays $\ell_\gamma$ become coincident.  These loci
are called \ti{walls of marginal stability}, because as we cross them the
$\Omega(\gamma)$ jump discontinuously (i.e. some BPS bound states appear or decay).
Since $\Omega(\gamma)$ appears in \eqref{eq:psiinst-pieces}-\eqref{eq:psiinst2},
we might worry that $\Psi^\sigma$ would jump discontinuously at these walls.
Were this to happen, it could mean that the holomorphic structure induced by $\Psi^\sigma$
would also jump discontinuously.
I believe that such discontinuities do not occur:  the holomorphic structure
is continuous across the walls of marginal stability.  Unfortunately, though, I will
not be able to give a complete demonstration of that fact in this note; below I will only
explain how it works in the simplest example and give some indication of how that
example ought to generalize.

\subsubsection{A simple example:  the pentagon}

Let us consider the most basic example of this phenomenon, to get some idea
of what to expect.  Suppose that at some point $u$ of $\cB$
we have two charges $\gamma_1$, $\gamma_2$,
with $\IP{\gamma_1, \gamma_2} = 1$, and $\Omega(\pm \gamma_1) = 1$, $\Omega(\pm \gamma_2) = 1$,
$\ell_{\gamma_2}$ is displaced slightly clockwise from $\ell_{\gamma_1}$,
and all other $\Omega(\gamma)$ vanish.
Let us use the notation $x$ and $y$ to denote holomorphic functions of $\zeta$, with
$(x,y) = (\cY_{\gamma_1},\cY_{\gamma_2})$ when evaluated slightly clockwise from  $\ell_{\gamma_2}$.
Importantly, this does not imply that $(x,y) = (\cY_{\gamma_1}, \cY_{\gamma_2})$
everywhere; the reason is that, as we have discussed, $\cY_{\gamma_1}$ is discontinuous along the ray $\ell_{\gamma_2}$.  Instead we have
\begin{equation}
 \cY_{\gamma_2} = y \text{ on } {\ell_{\gamma_2}}, \qquad \cY_{\gamma_1} = x-xy \text{ on }  {\ell_{\gamma_1}}.
\end{equation}
Now consider displacing $u$ to a point on the other side of a wall of marginal stability,
where $\ell_{\gamma_1}$ and $\ell_{\gamma_2}$ have exchanged positions.  At this point
one has $\Omega(\gamma_1) = \Omega(\gamma_2) = \Omega(\gamma_1 + \gamma_2) = 1$.  In consequence,
the discontinuity structure of $\cY$ is more complicated, and we have
\begin{equation}
 \cY_{\gamma_1} = x \text{ on }  {\ell_{\gamma_1}}, \qquad \cY_{\gamma_1 + \gamma_2} = - \frac{xy}{1-x} \text{ on } \ell_{\gamma_1 + \gamma_2}, \qquad \cY_{\gamma_2} = \frac{y}{1-x+xy} \text{ on } {\ell_{\gamma_2}}.
\end{equation}
Now consider the limit where $u$ approaches the wall (from either side).
In this limit all the rays collapse to a single $\ell$, and
\eqref{eq:psiinst1}, \eqref{eq:psiinst2}
become integrals over that $\ell$.  We would like to show that the integrands
are independent of which side of the wall we approach from.

For $\Psi^{\inst,\sigma,1}$ this is a consequence of the 5-term dilogarithm identity\footnote{To
be precise we have to specify the branch choices here.  They are fixed by the requirement
that we take the principal branch when all $\xi^\sigma_\gamma = 0$.  At this locus
we have $x, y \in \R_-$.}
\begin{equation} \label{eq:logint-rogers}
 \cR(y) + \cR(x-xy) = \cR(x) + \cR\left(- \frac{xy}{1-x}\right) + \cR \left(\frac{y}{1-x+xy}\right),
\end{equation}
where $\cR$ denotes the ``Rogers dilogarithm'' function
\begin{equation}
 \cR(z) = \Li_2(z) + \half \log(-z) \log(1-z).
\end{equation}
For $\Psi^{\inst,\sigma,2}$ it boils down to the more elementary identities
\begin{align}
 \log(1 - x + xy) &= \log(1 - x) + \log\left(1 + \frac{xy}{1-x}\right), \label{eq:logint-x} \\
 \log(1 - y) &= \log\left(1 + \frac{xy}{1-x}\right) + \log\left(1 - \frac{y}{1-x+xy}\right). \label{eq:logint-y}
\end{align}
(Note that \eqref{eq:logint-x}, \eqref{eq:logint-y} are closely related to \eqref{eq:logint-rogers} ---
indeed, they are the shifts of \eqref{eq:logint-rogers} under analytic continuation of $x$, $y$
respectively around $0$.  So in a certain sense \eqref{eq:logint-rogers} is all that we really needed.)
This completes the proof that $\Psi^\sigma$ is continuous in this case.

\subsubsection{More general examples} \label{eq:general-wc}

In a more general case, showing that $\Psi^\sigma$ is continuous across walls
of marginal stability would involve dilogarithm identities generalizing \eqref{eq:logint-rogers}.
From our present point of view, these
identities should be understood are consequences of the wall-crossing formula for the BPS spectrum.
More precisely, they should
follow from the ``refined'' version of the wall-crossing formula, first precisely
stated in \cite{Dimofte:2009tm}, and physically derived in \cite{Dimofte:2009tm,Cecotti:2009uf,Gaiotto:2010be}.

Let us first explain how this works for the identity \eqref{eq:logint-rogers}.
We consider a \ti{noncommutative} algebra with generators
$\hx$, $\hy$ subject to the relation
\begin{equation} \label{eq:xy-algebra}
 q^{-1} \hx\hy = q \hy\hx.
\end{equation}
Applying the \ti{quantum dilogarithm}
\begin{equation}
\bE(z) = \prod_{n = 0}^\infty (1 + q^{2n+1} z)^{-1}
\end{equation}
to these noncommutative variables, we have the identity \cite{Faddeev:1993rs}
\begin{equation} \label{eq:pentagon}
 \bE(\hx) \bE(\hy) = \bE(\hy) \bE(q^{-1} \hx\hy) \bE(\hx).
\end{equation}
This identity is an instance of the general refined wall-crossing formula of \cite{Dimofte:2009tm,Cecotti:2009uf,Gaiotto:2010be}.  On the left we see two BPS hypermultiplets
with charges $\gamma_1$ and $\gamma_2$; on the right we see three BPS hypermultiplets
with charges $\gamma_2$, $\gamma_1 + \gamma_2$, $\gamma_1$.
But on the other hand, it
was explained in \cite{Faddeev:1993rs} that \eqref{eq:pentagon} is a sort of quantization
of \eqref{eq:logint-rogers}, and in particular that one can recover \eqref{eq:logint-rogers}
from it.  The idea is to consider the algebra \eqref{eq:xy-algebra} as represented on
a Hilbert space and then consider the $q \to 1$ asymptotics of the
symbols of the operators on both sides of \eqref{eq:pentagon}.

It is natural to expect that the same approach can be adapted more generally,
beginning from the general refined wall-crossing formula and studying its $q \to 1$ limit
to give the necessary dilogarithm identities.  It is straightforward to see
that one will get an identity of roughly the correct form, in the sense
that one sees a sum of functions $\Li_2$ appearing with the correct arguments.
I have not carried out a verification that in this way one gets precisely the correct identities.

\section{Physical interpretation} \label{sec:physical-interpretation}

The bulk of this paper has been devoted to the description of a hyperholomorphic line bundle
$V$ over the target space $\cM$ of a 3-dimensional sigma model $T_3[R]$.  In this final section
I make a tentative proposal about the physical meaning of this line bundle.

The sigma model $T_3[R]$ arose by compactifying a 4-dimensional field theory $T_4$ on $S^1$ with radius $R$.
In particular, this sigma model is supposed to describe the infrared physics of $T_4$ on $X = S^1 \times \R^3$.
In order to understand $V$, we consider the next simplest possibility:  put
$T_4$ on $Y = \text{Taub-NUT space of radius $R$}$.  $Y$ has an $S^1$ acting by
isometries, and the space of orbits is $\R^3$.
The important difference between $X$ and $Y$ is that on $Y$ the $S^1$ action is not free:  there is a single
fixed point (called a ``NUT center''), which we may take to be the origin of $\R^3$.

In either case, let us study the physics at some length scale $L \gg R$.
At this scale the field theory is effectively 3-dimensional.  If we go out a distance $L' \gg L$ from the origin
of $\R^3$ and look at the local physics in regions of size $L$,
then it is very difficult to detect a difference between
$X$ and $Y$.  The difference between the two compactifications makes itself felt only when we
come close to the origin in $\R^3$.

In other words, the NUT center appears
to be a kind of local operator $\cO_{\NUT}$ which can be inserted in the
theory $T_3[R]$.
Since $T_3[R]$ is a sigma model of maps $\varphi: \R^3 \to \cM$,
we may ask:  how can $\cO_{\NUT}$ be described in the sigma model language?

The simplest possibility would be that there is some function $f_{\NUT}: \cM \to \C$
such that inserting $\cO_{\NUT}$ at a point $c \in \R^3$ corresponds to inserting
$f_\NUT(\varphi(c))$ in the sigma-model path integral.  For example, this is
what happens for the local operators $\cO_L$ of $T_3[R]$ that are obtained by wrapping line
operators $L$ of $T_4$ around $S^1$.

What I propose is that, rather than a function on $\cM$, $\cO_{\NUT}$ corresponds to a \ti{section} $s_{\NUT}$ of the line bundle $V$.
While I have no really direct argument, there are several pieces of circumstantial evidence for this proposal:

\begin{itemize}
 \item The first (worked out
in discussions with Dan Freed) exploits an observation made by Greg Moore:
if one carefully carries out the compactification of $T_4$ on the $S^1$ fiber of $Y$,
one finds that the reduced Lagrangian includes a
new interaction term of the form
\begin{equation}
 I_\CS = \frac{1}{2 \pi} \int_M A \wedge \varphi^*(F),
\end{equation}
where $A$ is the fixed $U(1)$ connection in the $S^1$ fiber of $Y$,
and $F$ is the curvature of a $U(1)$ connection in the line bundle $V$ over $\cM$.

The term $I_\CS$ is gauge invariant on closed 3-manifolds $M$, but not on $M$ with boundary.
In consequence, when $M$ has boundary, the exponentiated action $E(\varphi)$
is not a number but an element of a line
$L(\varphi)$.  The line $L(\varphi)$ depends on the restriction $\varphi \vert_{\partial M}$
as well as on $A \vert_{\partial M}$.
In particular, suppose $\partial M = S^2$, $A$ has
Chern class $k$ over $\partial M$, and $\varphi \vert_{\partial M}$ is a constant map,
with value say $x \in \cM$.  In this case one finds that
$L(\varphi)$ is simply the fiber of $V^{-k}$ at $x$.

To define the sigma-model action in the presence of a NUT
center $c \in \R^3$, we should bore out a small ball $B_c$
around $c$ of size $\epsilon$, and then take a limit as $\epsilon \to 0$.
The preimage of $B_c$ in $Y$ is a 4-ball.  The boundary $S^3$ of this 4-ball
is Hopf-fibered over a $\partial B_c \simeq S^2$.
Effectively, then, we are studying the theory on a manifold with
boundary $S^2$, over which the $U(1)$ connection has Chern class 1.
So from the above
we conclude that the exponentiated
sigma-model action $E(\varphi)$ is naturally valued in the fiber of
$V^{-1}$ over $\varphi(c)$.  The full integrand of the path integral includes
in addition the operator $\cO_\NUT(c)$; and this integrand
should be a number, not an element of a nontrivial line.
So in order that the sigma model path integral have a chance of making sense,
we should require that $\cO_\NUT$ is a section of $V$.

\item A second reason to expect that $\cO_\NUT$ is represented by
a section of $V$ comes from string dualities.
Suppose that the theory $T_4$ is realized by compactification of Type IIB string theory
on a non-compact Calabi-Yau threefold $Z$.
By $T$-duality along the circle fiber of the NUT space,
one can transform a NUT center into a Euclidean Type IIA NS5-brane wrapped on $Z$
\cite{Ooguri:1996wj}.  Lifting to M-theory we obtain a fivebrane on $Z$.
The partition function of the fivebrane has been studied in \cite{Witten:1997hc};
the result is that it is most naturally viewed as a section of a line bundle over
the intermediate Jacobian of $Z$.  The topology of this line bundle
matches that of $V$, at least when restricted to the individual torus fibers of $\cM$.

\end{itemize}

Now, what is the physical significance of the fact that $V$ is a \ti{hyperholomorphic} bundle?
We have noted earlier that the complex structures $J_\zeta$ on $\cM$ correspond to subalgebras
$A_\zeta$ of the supersymmetry algebra.
A NUT center also preserves 1/2 of the supersymmetry algebra:  indeed the subalgebra it
preserves is simply $A_{\zeta = 0}$.
Thus it is natural to expect that $s_{\NUT}$ should be holomorphic when considered
as a section of $V_{\zeta = 0}$ over $\cM_{\zeta = 0}$.  So at least \ti{one} of the holomorphic
structures on $V$ has a natural role to play.

What about the other holomorphic structures?
There is a natural generalization of the above:
we may consider an $\Omega$-deformed version of the theory $T_4$
on $Y$ \cite{Nekrasov:2002qd,Nekrasov:2003rj}, using the $U(1)$ isometry of $Y$.
This deformation depends on a parameter $\varepsilon$ with dimensions of mass.
As observed in \cite{Nekrasov:2010ka}, there is a change of variable which
shows that the $\Omega$-deformed theory is locally equivalent to the original one, wherever the $U(1)$ action is free --- in other words,
everywhere except at the NUT center.
The equivalence however
maps $A_{\zeta = 0}$ to some other $A_\zeta$, where $\zeta$ depends on the dimensionless combination $\varepsilon R$.

Hence, if we begin with the $\Omega$-deformed theory on $Y$
and make this change of variable, we will end up with the theory $T_3[R]$
away from the origin of $\R^3$, with a 1/2-BPS local operator $\cO_{\NUT}(\zeta)$
inserted at the origin, represented by a corresponding section $s_\NUT(\zeta)$ of $V$.
Since $\cO_\NUT(\zeta)$ preserves the algebra $A_\zeta$,
$s_\NUT(\zeta)$ should be holomorphic when considered as a section of
$V_\zeta$ over $\cM_\zeta$.  We have thus found a natural role for the other
holomorphic structures on $V$ (although not for the single unitary connection $D$
which induces all of them.)

\bigskip

A consequence of this proposal is that not only does $\cM$ have a canonical hyperholomorphic line
bundle $V$, this line bundle even
has a canonical holomorphic section $s_{\NUT}(\zeta)$, which controls the IR physics
of NUT centers in circle compactifications of $T_4$.  It would be very interesting to
determine $s_{\NUT}(\zeta)$.

\section*{Acknowledgements}

I especially thank Boris Pioline for
our collaboration in 2008 (as noted in the introduction, this work really had its genesis in
that collaboration), as well as for more recent discussions.
I also thank Sergei Alexandrov, Dan Freed, Davide Gaiotto, Tamas Hausel and Greg Moore for
useful discussions.

This work was supported by NSF grant DMS-1006046.  I thank the Kavli Institute for
Theoretical Physics for hospitality while this paper was being completed.

\bibliography{hyperhol-paper}

\providecommand{\href}[2]{#2}\begingroup\raggedright\begin{thebibliography}{10}

\bibitem{AlvarezGaume:1981hm}
L.~Alvarez-Gaume and D.~Z. Freedman, ``{Geometrical Structure and Ultraviolet
  Finiteness in the Supersymmetric Sigma Model},'' {\em Commun. Math. Phys.}
  {\bf 80} (1981)
443.

\bibitem{Seiberg:1996nz}
N.~Seiberg and E.~Witten, ``Gauge dynamics and compactification to three
  dimensions,''
\href{http://www.arXiv.org/abs/hep-th/9607163}{{\tt hep-th/9607163}}.

\bibitem{Ooguri:1996me}
H.~Ooguri and C.~Vafa, ``{S}umming up {D}-instantons,'' {\em Phys. Rev. Lett.}
  {\bf 77} (1996) 3296--3298,
\href{http://www.arXiv.org/abs/hep-th/9608079}{{\tt hep-th/9608079}}.

\bibitem{Seiberg:1996ns}
N.~Seiberg and S.~H. Shenker, ``Hypermultiplet moduli space and string
  compactification to three dimensions,'' {\em Phys. Lett.} {\bf B388} (1996)
  521--523,
\href{http://www.arXiv.org/abs/hep-th/9608086}{{\tt hep-th/9608086}}.

\bibitem{Gaiotto:2008cd}
D.~Gaiotto, G.~W. Moore, and A.~Neitzke, ``{Four-dimensional wall-crossing via
  three-dimensional field theory},'' {\em Commun. Math. Phys.} {\bf 299} (2010)
  163--224,
\href{http://www.arXiv.org/abs/0807.4723}{{\tt 0807.4723}}.

\bibitem{Gaiotto:2010be}
D.~Gaiotto, G.~W. Moore, and A.~Neitzke, ``{Framed BPS States},''
\href{http://www.arXiv.org/abs/1006.0146}{{\tt 1006.0146}}.

\bibitem{Kapustin:2006hi}
A.~Kapustin, ``{Holomorphic reduction of N = 2 gauge theories, Wilson-'t Hooft
  operators, and S-duality},''
\href{http://www.arXiv.org/abs/hep-th/0612119}{{\tt hep-th/0612119}}.

\bibitem{Kapustin:2007wm}
A.~Kapustin and N.~Saulina, ``{The algebra of Wilson-'t Hooft operators},''
  {\em Nucl. Phys.} {\bf B814} (2009) 327--365,
\href{http://www.arXiv.org/abs/0710.2097}{{\tt 0710.2097}}.

\bibitem{Alday:2009aq}
L.~F. Alday, D.~Gaiotto, and Y.~Tachikawa, ``{Liouville Correlation Functions
  from Four-dimensional Gauge Theories},'' {\em Lett. Math. Phys.} {\bf 91}
  (2010) 167--197,
\href{http://www.arXiv.org/abs/0906.3219}{{\tt 0906.3219}}.

\bibitem{Alday:2009fs}
L.~F. Alday, D.~Gaiotto, S.~Gukov, Y.~Tachikawa, and H.~Verlinde, ``{Loop and
  surface operators in N=2 gauge theory and Liouville modular geometry},''
\href{http://www.arXiv.org/abs/0909.0945}{{\tt 0909.0945}}.

\bibitem{MR2233852}
V.~Fock and A.~Goncharov, ``Moduli spaces of local systems and higher
  {T}eichm\"uller theory,'' {\em Publ. Math. Inst. Hautes \'Etudes Sci.}
  (2006), no.~103, 1--211, \href{http://www.arXiv.org/abs/math/0311149}{{\tt
  math/0311149}}.

\bibitem{MR2567745}
V.~V. Fock and A.~B. Goncharov, ``Cluster ensembles, quantization and the
  dilogarithm,'' {\em Ann. Sci. \'Ec. Norm. Sup\'er. (4)} {\bf 42} (2009),
  no.~6, 865--930, \href{http://www.arXiv.org/abs/math/0311245}{{\tt
  math/0311245}}.

\bibitem{Gaiotto:2009fs}
D.~Gaiotto, ``{Surface Operators in N=2 4d Gauge Theories},''
\href{http://www.arXiv.org/abs/0911.1316}{{\tt 0911.1316}}.

\bibitem{Taki:2010bj}
M.~Taki, ``{Surface Operator, Bubbling Calabi-Yau and AGT Relation},''
\href{http://www.arXiv.org/abs/1007.2524}{{\tt 1007.2524}}.

\bibitem{Awata:2010bz}
H.~Awata, H.~Fuji, H.~Kanno, M.~Manabe, and Y.~Yamada, ``{Localization with a
  Surface Operator, Irregular Conformal Blocks and Open Topological String},''
\href{http://www.arXiv.org/abs/1008.0574}{{\tt 1008.0574}}.

\bibitem{Gaiotto:2011tf}
D.~Gaiotto, G.~W. Moore, and A.~Neitzke, ``{Wall-Crossing in Coupled 2d-4d
  Systems},''
\href{http://www.arXiv.org/abs/1103.2598}{{\tt 1103.2598}}.

\bibitem{Alexandrov:2010np}
S.~Alexandrov, D.~Persson, and B.~Pioline, ``{On the topology of the
  hypermultiplet moduli space in type II/CY string vacua},'' {\em Phys.Rev.}
  {\bf D83} (2011) 026001, \href{http://www.arXiv.org/abs/1009.3026}{{\tt
  1009.3026}}.

\bibitem{Ward:1977ta}
R.~S. Ward, ``{O}n selfdual gauge fields,'' {\em Phys. Lett.} {\bf A61} (1977)
81--82.

\bibitem{Hitchin:1986ea}
N.~J. Hitchin, A.~Karlhede, U.~Lindstrom, and M.~Ro\v{c}ek, ``Hyperk{\"a}hler
  metrics and supersymmetry,'' {\em Commun. Math. Phys.} {\bf 108} (1987)
535.

\bibitem{MR1919716}
B.~Feix, ``Hypercomplex manifolds and hyperholomorphic bundles,'' {\em Math.
  Proc. Cambridge Philos. Soc.} {\bf 133} (2002), no.~3, 443--457.

\bibitem{Cecotti:1993rm}
S.~Cecotti and C.~Vafa, ``On classification of {$\N=2$} supersymmetric
  theories,'' {\em Commun. Math. Phys.} {\bf 158} (1993) 569--644,
\href{http://www.arXiv.org/abs/hep-th/9211097}{{\tt hep-th/9211097}}.

\bibitem{Seiberg:1994rs}
N.~Seiberg and E.~Witten, ``Electric-magnetic duality, monopole condensation,
  and confinement in {$\N=2$} supersymmetric {Y}ang-{M}ills theory,'' {\em
  Nucl. Phys.} {\bf B426} (1994) 19--52,
\href{http://www.arXiv.org/abs/hep-th/9407087}{{\tt hep-th/9407087}}.

\bibitem{Denef:2007vg}
F.~Denef and G.~W. Moore, ``Split states, entropy enigmas, holes and halos,''
\href{http://www.arXiv.org/abs/hep-th/0702146}{{\tt hep-th/0702146}}.

\bibitem{ks1}
M.~Kontsevich and Y.~Soibelman, ``{S}tability structures, motivic
  {D}onaldson-{T}homas invariants and cluster transformations,''
  \href{http://www.arXiv.org/abs/0811.2435}{{\tt 0811.2435}}.

\bibitem{Cecotti:2009uf}
S.~Cecotti and C.~Vafa, ``{BPS Wall Crossing and Topological Strings},''
\href{http://www.arXiv.org/abs/0910.2615}{{\tt 0910.2615}}.

\bibitem{Cecotti:2010qn}
S.~Cecotti and C.~Vafa, ``{2d Wall-Crossing, R-twisting, and a Supersymmetric
  Index},''
\href{http://www.arXiv.org/abs/1002.3638}{{\tt 1002.3638}}.

\bibitem{Manschot:2010qz}
J.~Manschot, B.~Pioline, and A.~Sen, ``{Wall-Crossing from Boltzmann Black Hole
  Halos},'' {\em JHEP} {\bf 07} (2011) 059,
\href{http://www.arXiv.org/abs/1011.1258}{{\tt 1011.1258}}.

\bibitem{Dimofte:2009tm}
T.~Dimofte, S.~Gukov, and Y.~Soibelman, ``{Quantum Wall Crossing in N=2 Gauge
  Theories},''
\href{http://www.arXiv.org/abs/0912.1346}{{\tt 0912.1346}}.

\bibitem{Alexandrov:2011ac}
S.~Alexandrov, D.~Persson, and B.~Pioline, ``{Wall-crossing, Rogers
  dilogarithm, and the QK/HK correspondence},''
\href{http://www.arXiv.org/abs/1110.0466}{{\tt 1110.0466}}.

\bibitem{MR2394039}
A.~Haydys, ``Hyper{K}\"ahler and quaternionic {K}\"ahler manifolds with
  {$S^1$}-symmetries,'' {\em J. Geom. Phys.} {\bf 58} (2008), no.~3, 293--306,
  \href{http://www.arXiv.org/abs/0706.4473}{{\tt 0706.4473}}.

\bibitem{hitchin-haydys-talk}
N.~Hitchin, ``{Higgs bundles and quaternionic geometry},'' 2011.
\newblock Talk at the Isaac Newton Institute for Mathematical Sciences,
  available at {\tt
  http://www.newton.ac.uk/programmes/MOS/seminars/070111301.html}.

\bibitem{Seiberg:1994aj}
N.~Seiberg and E.~Witten, ``Monopoles, duality and chiral symmetry breaking in
  {$\N=2$} supersymmetric {Q}{C}{D},'' {\em Nucl. Phys.} {\bf B431} (1994)
  484--550,
\href{http://www.arXiv.org/abs/hep-th/9408099}{{\tt hep-th/9408099}}.

\bibitem{Gaiotto:2009hg}
D.~Gaiotto, G.~W. Moore, and A.~Neitzke, ``{Wall-crossing, Hitchin Systems, and
  the WKB Approximation},''
\href{http://www.arXiv.org/abs/0907.3987}{{\tt 0907.3987}}.

\bibitem{Cecotti:1989qn}
S.~Cecotti, S.~Ferrara, and L.~Girardello, ``Geometry of type {I}{I}
  superstrings and the moduli of superconformal field theories,'' {\em Int. J.
  Mod. Phys.} {\bf A4} (1989)
2475.

\bibitem{Faddeev:1993rs}
L.~D. Faddeev and R.~M. Kashaev, ``Quantum dilogarithm,'' {\em Mod. Phys.
  Lett.} {\bf A9} (1994) 427--434,
\href{http://www.arXiv.org/abs/hep-th/9310070}{{\tt hep-th/9310070}}.

\bibitem{Ooguri:1996wj}
H.~Ooguri and C.~Vafa, ``Two-dimensional black hole and singularities of {CY}
  manifolds,'' {\em Nucl. Phys.} {\bf B463} (1996) 55--72,
\href{http://www.arXiv.org/abs/hep-th/9511164}{{\tt hep-th/9511164}}.

\bibitem{Witten:1997hc}
E.~Witten, ``{F}ive-brane effective action in {M}-theory,'' {\em J. Geom.
  Phys.} {\bf 22} (1997) 103--133,
\href{http://www.arXiv.org/abs/hep-th/9610234}{{\tt hep-th/9610234}}.

\bibitem{Nekrasov:2002qd}
N.~A. Nekrasov, ``{S}eiberg-{W}itten prepotential from instanton counting,''
  {\em Adv. Theor. Math. Phys.} {\bf 7} (2004) 831--864,
\href{http://www.arXiv.org/abs/hep-th/0206161}{{\tt hep-th/0206161}}.

\bibitem{Nekrasov:2003rj}
N.~A. Nekrasov and A.~Okounkov, ``{S}eiberg-{W}itten theory and random
  partitions,''
\href{http://www.arXiv.org/abs/hep-th/0306238}{{\tt hep-th/0306238}}.

\bibitem{Nekrasov:2010ka}
N.~Nekrasov and E.~Witten, ``{The Omega Deformation, Branes, Integrability, and
  Liouville Theory},''
\href{http://www.arXiv.org/abs/1002.0888}{{\tt 1002.0888}}.

\end{thebibliography}\endgroup

\small\normalsize

\end{document}